\documentclass[prc,floatfix,nofootinbib,superscriptaddress,showpacs,preprintnumbers,aps]{revtex4}
\usepackage {amsfonts}
\usepackage {graphicx}
\usepackage {color}
\begin{document}
\preprint{PRC/Draft \today}
\title{Rapid fluctuation of the tensor polarization of deuteron beams behind carbon foils and their relation to the resonances in the isospin-breaking  $^{12}{\rm C}(d,\alpha_{2})^{10}{\rm B}^{*}$ reaction}
\author{H.~Seyfarth}
\email[E-mail: ]{H.Seyfarth@fz-juelich.de}
\affiliation{Institut f\"ur Kernphysik, J\"ulich Center for Hadron Physics, Forschungszentrum J\"ulich, Leo--Brandt--Str.\ 1, D-52425 J\"ulich, Germany}
\author{A.~Rouba}
\author{V.~Baryshevsky}
\affiliation{Research Institute for Nuclear Problems, Bobruiskaya Str.\ 11, 220050 Minsk, Belarus}
\author{R.~Engels}
\author{K.~Grigoryev}
\author{F.~Rathmann}
\author{H.~Str\"oher}
\affiliation{Institut f\"ur Kernphysik, J\"ulich Center for Hadron Physics, Forschungszentrum J\"ulich, Leo--Brandt--Str.\ 1, D-52425 J\"ulich, Germany}
\date{\today}
\begin{abstract}
Rapid fluctuations are observed in the tensor-polarization $p_{zz}$ of deuteron beams forward-transmitted through graphite targets. Unpolarized 9.50 to 18.60\,MeV beams from the K\"oln tandem accelerator were utilized, and the polarization behind seven 36 to 188\,mg/cm$^{2}$ targets was measured with a polarimeter based on the $^{3}{\rm He}(\vec{d},p)^{4}{\rm He}$ reaction. Due to the chosen relation between the areal target densities and the initial beam energies $E_{\rm in}$, the seven sets of $p_{zz}(E_{\rm in})$ can be combined in a common plot as a function of $E_{\rm in}$. This allows one to understand $p_{zz}$, measured behind the 188\,mg/cm$^{2}$ target at  $E_{\rm in}$=18.6\,MeV, as resulting from the sequence of differential polarization production $\Delta p_{zz}(E)/\Delta E$ during energy degradation in the target from $E$=18.60 to 9.50\,MeV. The rapid fluctuations of $\Delta p_{zz}(E)/\Delta E$$p_{zz}$ are described by 51 Gaussian-distributed cross-sections removing deuterons either in the $m=0$ or in the $m=\pm 1$ state from the beam. The 51 fitted central energies $E_{0}$  with a single exception agree with the energies of the narrow peaks in the excitation functions of the weak, isospin-breaking $^{12}{\rm C}(d,\alpha_{2})^{10}{\rm B}^{*}(1.74\,{\rm MeV},J^{\pi}=0^{+},T=1)$ reaction with population of the second excited $^{10}{\rm B}$ state via intermediate excited $^{14}$N states. Angular momentum and parity conservation confines their spin values to $J=l$, the orbital angular momentum in the entrance state. Formation of $J=l\pm 1$ states with other decay channels would be possible. According to the fit result, formation of $J=l$ excited $^{14}$N states dominates in the removal of deuterons from the beam. The coarse energy dependence of the present fluctuations corresponds to the excitation of the $^{14}$N giant electric dipole resonance, split into regions of $\Delta p_{zz}(E)/\Delta E$ $>$0 and $<$0. Strong evidence is found that removal of deuterons in the $m=0$ ($m=\pm 1$) state from the beam leads to the formation of $^{14}$N states of established positive (negative) parity. As an application, the removal cross-section functions allow to calculate $p_{zz}$ achievable with carbon targets for initial beam energies $E_{\rm in}$ between 18.60 and 9.50\,MeV and $E_{\rm out}$ given by the areal target density. Carbon layers in a sandwich technique would enable $p_{zz}$ between -0.4 and +0.3.
\end{abstract}
\maketitle
\section{Introduction~\label{Introduction}}
The appearance of tensor polarization in initially unpolarized deuteron beams, forward-transmitted through foils of spin-zero nuclei, is expected as consequence of the quadrupole deformation of the deuteron, an effect called nuclear dichroism in analogy to the optical effect~\,\cite{Baryshevsky_1992,Baryshevsky_1993,Baryshevsky_2012}. The deviation from spherical shape leads to a difference of the deuteron-nucleus cross sections $\sigma_{0}$ and $\sigma_{\pm1}$ for deuterons (spin 1) in the $m=0$ and $m=\pm1$ state, respectively. The quantum numbers are defined relative to the direction of the initial beam as the only available quantization axis. The quantization axis of the forward-transmitted beam coincides with that of the initial beam. Due to the azimuthal symmetry around the quantization axis and parity conservation in the nuclear interaction, the polarization of the forward-transmitted beam is fully described by the tensor-polarization component $p_{zz}$ (e.g., \cite{Ohlsen_1972}). Because of the cigar-shaped quadrupole deformation of the deuteron, the cross section $\sigma_{0}$ with the deuteron deformation axis and spin perpendicular to the beam axis (m=0) is larger than $\sigma_{\pm1}$ with orientation along the axis ($m=\pm1$). Defined by the the difference of the relative occupation of the $m=0$ and $m=\pm1$ states in the transmitted beam, a positive sign of $p_{zz}$ is expected for $\sigma_{0}>\sigma_{\pm1}$. According to the earlier theoretical papers\,\cite{Baryshevsky_1992,Baryshevsky_1993} one expected $p_{zz}$ of the order of +0.01 for the deuteron beam behind a 100\,mg/cm$^{2}$ carbon target  (predominantly spin-zero $^{12}$C) and initial beam energies around 10\,MeV. Furthermore, a smooth dependence of $p_{\rm zz}$ on the target thickness was to be expected.

These predictions were studied in two beam times (in 2003 and 2006) with unpolarized deuteron beams from the HVEC FN Van-de-Graaf tandem accelerator of the Institut f\"ur Kernphysik of  Universit\"at zu K\"oln\,\cite{Jolie_2002}, seven graphite targets of different areal densities, and polarimetry of the forward-transmitted deuteron beam with use of the $\vec{d}+{^{3}\rm{He}}\to p+{^{4}\rm{He}}$ reaction\,\cite{Engels_1997}. Reference data were measured without target and with two gold foils of different areal densities. At a beam energy $E_{\rm c}$ in the polarimeter $^3$He gas cell, $p_{zz}(E_{\rm c})$ is deduced from the ratio of proton counts in the four side detectors and in the forward detector, $r^{\rm Cx}(E_{\rm c})$ in the measurements with a carbon target (Cx) and $r^{\rm ref}$ at the same energy from the reference measurement. Since the beam energies in the polarimeter cell in the reference measurements do not coincide with those in the measurements with the graphite targets, the values $r^{\rm ref}(E_{\rm c})$ have to be taken from a fit to the set of measured ratios. In our first analysis\,\cite{Seyfarth_2010}, in expectation of a weak and smooth energy dependence for each graphite target, linear fits were applied to the set of ratios $r^{\rm Cx}(E_{\rm c})$, too. Accordingly, for each of the seven target a smooth dependence of $p_{zz}$ on the energy $E_{\rm c}$ was obtained. The strong variations, most prominent in the change of $p_{zz}$ from $-0.28$ at $E_{\rm in}$=14.8\,MeV to $-0.06$ at $E_{\rm in}$=15.6\,MeV measured with the 129.49\,mg/cm$^{2}$ target, are in evident contradiction to the theoretically predicted small positive value.

For relativistic energies the effect of nuclear dichroism was studied with the use of a 5.5\,GeV deuteron beam from the Nuclotron at Joint Institute for Nuclear Research Dubna and transmission through 40, 83, and 123\,g/cm$^{2}$ carbon targets\,\cite{Azhgirey_2008_1}. The polarization $p_{zz}$ of the transmitted beam was determined by scattering on a 10\,cm beryllium target and utilizing its high tensor analyzing power. The three measured values of $p_{zz}$ were fitted by the relation
\begin{equation}
p_{zz}
    =\frac{\displaystyle 1-{\rm e}^{-Nx\Delta\sigma} }
   {\displaystyle 1+\frac{1}{2}{\rm e}^{-Nx\Delta\sigma}},
\label{Pzz_Dubna}
\end{equation}
where $N$ (atoms/g) is the atom number density of the carbon targets and $x$ (g/cm$^{2}$) is the areal density. The fit parameter  $\Delta\sigma$ is the above mentioned cross-section difference $\Delta\sigma=\sigma_{0}-\sigma_{\pm 1}$. It was assumed not to depend on the energy degradation in the targets. The measurement yields a positive $p_{zz}$, increasing with the areal target density in agreement with the theoretical prediction. A fit to the three given $p_{zz}$ values yields $\Delta\sigma=(6.4\pm 1.4)\,{\rm fm}^{2}$, which is slightly larger than the theoretical values 3.87\,fm$^{2}$\,\cite{Azhgirey_2008_1} and 3.79\,fm$^{2}$\,\cite{Azhgirey_2008_2}, but appreciably larger than 1.3\,fm$^{2}$ resulting from the cross-sections given in an earlier calculation\,\cite{Faeldt_1980}.

In an attempt to understand the strong energy dependence of $p_{zz}$ around $E_{\rm in}$ of 14 to 15\,MeV, the interference in the nuclear and Coulomb interaction was taken into account in addition to the nuclear and Coulomb interactions\,\cite{Baryshevsky+Rouba_2010}. Inclusion of this effect yields for the deuteron-carbon interaction in the energy range 5 - 20\,MeV yields a sign change of $\sigma_{0}-\sigma_{\pm1}$ and therewith of the $p_{zz}$ production at about 11\,MeV. Deuterons, slowed down from 20 to 11\,MeV in a 180\,mg/cm$^2$ carbon target, would acquire $p_{zz}=+0.014$, whereas slowing down from 11 to 5.5\,MeV in a 70\,mg/cm$^2$ carbon target would yield $p_{zz}=-0.0035$. These calculations, too, failed to explain the measured energy dependence of $p_{zz}$. As a further step, the importance of the interaction of the electric quadrupole moment of the deuteron with the electric field of the target nucleus at short distances was studied in a recent work\,\cite{Baryshevsky+Rouba_2016}.

Rapid fluctuations of $r^{\rm Cx}$ as a function of $E_{\rm c}$ are found for all targets, smoothed by the fits in the earlier evaluation\,\cite{Seyfarth_2010}. These rapid variations are maintained in the present work by the use of the experimental ratios $r^{\rm Cx}(E_{\rm c})$ instead of the values from the fit functions. The theoretically predicted weak, smooth effect of nuclear dichroism gets superposed by these rapid fluctuations. Such fluctuations  were also observed in the elastic scattering of polarized deuterons on $^{12}$C in the energy range 20 to 30\,MeV\,\cite{Perrin_1972,Arvieux_1974}. It was mentioned there that inconsistencies in the description of the measured vector analyzing powers in the framework of the optical potential\,\cite{Satchler_1966} might be due to the creation of intermediate excited states in $^{14}$N. The angular distributions of low-energy deuteron induced reactions on $^{12}$C with beam energies of 2.502 and 2.735\,MeV had also been interpreted with creation of intermediate excited states in $^{14}$N\,\cite{McEllistrem_1958}. Due to the relation between the analyzing power and the polarization in the inverse reaction\,\cite{Haeberli_1974}, the fluctuations observed in the analyzing power are to be expected in the polarization of an unpolarized beam, too. Polarization production in initially unpolarized deuterons beams is a well known process for a long while\,\cite{Lakin_1955}. Deuteron scattering and stripping reactions were utilized to produce polarized beams before the development of the polarized atomic sources and the finding that electron stripping and acceleration of the ions can be done without essential loss of polarization. The polarizing effect, studied in the present work, is different as it does not concern the polarization of the interaction products but the polarization of the remaining, forward-transmitted deuteron beam. Both effects, however, are based on the difference of the interaction cross sections for the two deuteron spin orientations relative to a quantization axis. 

The present paper presents a more detailed description of the experimental procedure and the data analysis than it is found in our first paper\,\cite{Seyfarth_2010}. Section\,\ref{Theory} presents the formalism used to describe the production of $p_{zz}$ in the targets. Details of the experiment and the data processing are found in Sec.\,\ref{Setupetc}. Therein, Sec.\,\ref{Polarimetry} presents the experimental setup and the method to derive the $p_{zz}$ from the proton counts in the polarimeter detectors. The choice of targets and deuteron-beam energies is described in Sec.\,\ref{Targets+Beams}, the properties of the target foils are discussed in Sec.\,\ref{Prop_Foils}. The measurements and the results are presented in Sec.\,\ref{Meas+Res}. There, Sec.\,\ref{RefData} concerns those performed to obtain reference data with two gold foils and without a target. The measurements with the carbon targets and the resulting set of $p_{zz}(E_{\rm in})$ are found in Sec.\,\ref{Carbon}. The variety of uncertainties in the fit results is discussed in Sec.\,\ref{Uncertainties}. The modeling of the measured energy dependence of $p_{zz}$ with use of the formulae of Sec.\,\ref{Theory} and the results are presented in Sec.\,\ref{Modeling}. In Sec.\,\ref{Discussion} the observed fluctuations and the fit results are discussed , the coarse structure in Sec.\,\ref{Coarse structure} and the fine structure in Secs.\,\ref{Fine structure} and\,\ref{Signs}. The possibility to produce deuteron beams of positive or negative tensor polarization is described in Sec.\,\ref{Application}. Summarizing remarks are presented in Sec.\,\ref{Summary}.
\section{Production of tensor polarization $p_{zz}$ in a deuteron beam~\label{Theory}}
The initially unpolarized beam of spin-1 deuterons is characterized by equal population of the $m=+1, m=0$, and  $m=-1$ states.  The only available quantization axis ($z$) is along the beam direction and the fractional beam intensities are equal,  $I_{+1}^{0}=I_{0}^{0}=I_{-1}^{0}=1/3\cdot I^{0}$, where $I^{0}$ is the total intensity of the initial beam. The tensor-polarization component $p_{zz}$ can be written as (e.g.,\,\cite{Haeberli_1967})
\begin{equation}
 p_{zz}^{0}=\frac{I_{+1}^{0}+I_{-1}^{0}-2I_{0}^{0}}{I_{+1}^{0}+I_{0}^{0}+I_{-1}^{0}}
                        =\frac{I_{+1}^{0}+I_{-1}^{0}-2I_{0}^{0}}{I^{0}}.
\label{DefPzz}
\end{equation}
As the other polarization components it is zero. The direction of the quantization axis of the beam, forward-transmitted through a target, coincides with that of the initial beam. Non-vanishing values of $p_{zz}$ of the beam behind the target arise, when the interaction with the target nuclei leads to a difference between $I_{+1}+I_{-1}$ and $2\cdot I_{0}$. The effect on the fractional beam intensities can be described by energy-dependent cross sections $\sigma_{+1}(E)$, $\sigma_{0}(E)$, and $\sigma_{-1}(E)$, removing deuterons of energy $E$ in the $m=+1, m=0$ or $m=-1$ state from the beam. These removal cross sections include those for all reactions between the deuterons and the target nuclei including he cross section for deuteron scattering except that part which corresponds to forward scattering into the polarimeter cell. For spin-0 target nuclei and due to the symmetry in the wave function of the quadrupole-deformed deuteron one can assume $\sigma_{+1}(E)=\sigma_{-1}(E)=\sigma_{\pm 1}(E)$. Behind a thin target of atomic number density $\rho$, thickness $d$, and negligible energy loss, Eq.\,(\ref{DefPzz}) with $I_{m}(\rho d)=I^{0}_{m}\cdot \rm{exp}\{-\rho d \sigma_{\rm m}(\it{E})\}$ yields
\begin{eqnarray}
p_{zz}(\rho d)=p_{zz}(E)=\frac{2 \cdot {\rm exp}~\{-\rho d \sigma_{\pm 1}(E)\}-2 \cdot {\rm exp}~\{-\rho d \sigma_{0}(E)\}}
                                    {2 \cdot {\rm exp}~\{-\rho d \sigma_{\pm 1}(E)\}+{\rm exp}~\{-\rho d \sigma_{0}(E)\}},
\label{PzzThin1}
\end{eqnarray}
where $\rho$, $d$, and $\sigma$ are given in units of 1/volume, length, and area, respectively. When both $\rho d \sigma_{\pm1}(E)$ and $\rho d \sigma_{0}(E)$ are $\ll 1$, 
\begin{equation}
p_{zz}(E)=\frac{2}{3}\rho d [\sigma_{0}(E)-\sigma_{\pm1}(E)].
\label{PzzThin2}
\end{equation}
This relation is essential for the for the subsequent analysis: positive $p_{zz}$ is produced with $\sigma_{0}(E)>\sigma_{\pm 1}(E)$, negative $p_{zz}$ with $\sigma_{\pm1}(E)>\sigma_{0}(E)$.

In a thick target the deuterons are slowed down, the beam energy gets degraded. This is  taken into account by extending Eq.\,(\ref{PzzThin1}) to
\begin{equation}
p_{zz}(\rho d)=
        \frac{2\cdot {\rm exp}~\{-\rho \int_0^d\sigma_{\pm1}[E(x)]dx\} -2\cdot {\rm exp}~\{-\rho \int_0^d\sigma_{0}[E(x)]dx\}}
               {2\cdot {\rm exp}~\{-\rho \int_0^d\sigma_{\pm1}[E(x)]dx\}+{\rm exp}~\{-\rho \int_0^d\sigma_{0}[E(x)]dx\}},
\label{PzzThick1}
\end{equation}
where $\sigma_{\pm1}[E(x)]$ and $\sigma_{0}[E(x)]$ are the cross sections for the deuteron energy $E$ at a penetration depth $x$ into the target. The polarization of the beam behind the target, $p_{zz}(\rho d)$, results from the integral effect of the energy-dependent difference between $\sigma_{\pm1}[E(x)]$ and $\sigma_{0}[E(x)]$ during deceleration in the target of thickness $d$ (given in units of length) from $E(x=0)=E_{\rm in}$ to $E(x=d)=E_{\rm out}$.

When N$_{\pm 1}$ and N$_{0}$ energy-dependent cross sections contribute to the interaction, Eq.\,(\ref{PzzThick1}) has to be extended, 
\begin{equation}
 p_{zz}(\rho d)=
         \frac{2\cdot {\rm exp}~\{-\rho \int_0^d \sum_{{\rm i}=1}^{{\rm N}_{\pm 1}} \sigma_{\pm1,{\rm i}}[E(x)]dx\}
                -2\cdot {\rm exp}~\{-\rho \int_0^d\sum_{{\rm j}=1}^{{\rm N}_{0}}\sigma_{0,{\rm j}}[E(x)]dx\}}
                {2\cdot {\rm exp}~\{-\rho \int_0^d \sum_{{\rm i}=1}^{{\rm N}_{\pm 1}} \sigma_{\pm1,{\rm i}}[E(x)]dx\}
                 +{\rm exp}~\{-\rho \int_0^d \sum_{{\rm j}=1}^{{\rm N}_{0}} \sigma_{0,{\rm j}}[E(x)]dx\}}.
\label{PzzThick2}
\end{equation}
In the present work, appropriate choice of the energy-dependent cross sections in Eq.\,(\ref{PzzThick2}) is used to model the tensor polarizations measured behind the carbon targets.
\section{Experimental setup and data processing~\label{Setupetc}}
The experiment was carried out in two separate beam times in 2003 and 2006 at the HVEC FN Van-de-Graaff tandem accelerator of the Institut f\"ur Kernphysik of Universit\"at zu K\"oln\,\cite{Jolie_2002} with unpolarized deuteron beams from a sputter source covering the energy range 9.5 to 18.6\,MeV in the measurements with the carbon targets and 6.0 to 7.9\,MeV in those to get unpolarized reference data. The targets and the  polarimeter were installed in the beam line L15 behind the momentum-analyzing dipole magnet and the beam-switching magnet.

According to the settings of the slits behind the analyzing dipole magnet, defining the energy spread of the beam, and the diaphragms behind the last quadrupole magnet, the beam  diameter at the target position was expected to be $\le$2\,mm. This was confirmed by inspection of target foils after exposure to the beam. A slight change of the target-surface colour, caused by the beam, allowed one to determine the beam diameter at the target as $\sim$1.5\,mm.
\subsection{Polarimeter and procedure to derive $p_{zz}$~\label{Polarimetry}}
\begin{figure}
\begin{center}
\includegraphics[angle=90,width=8cm]{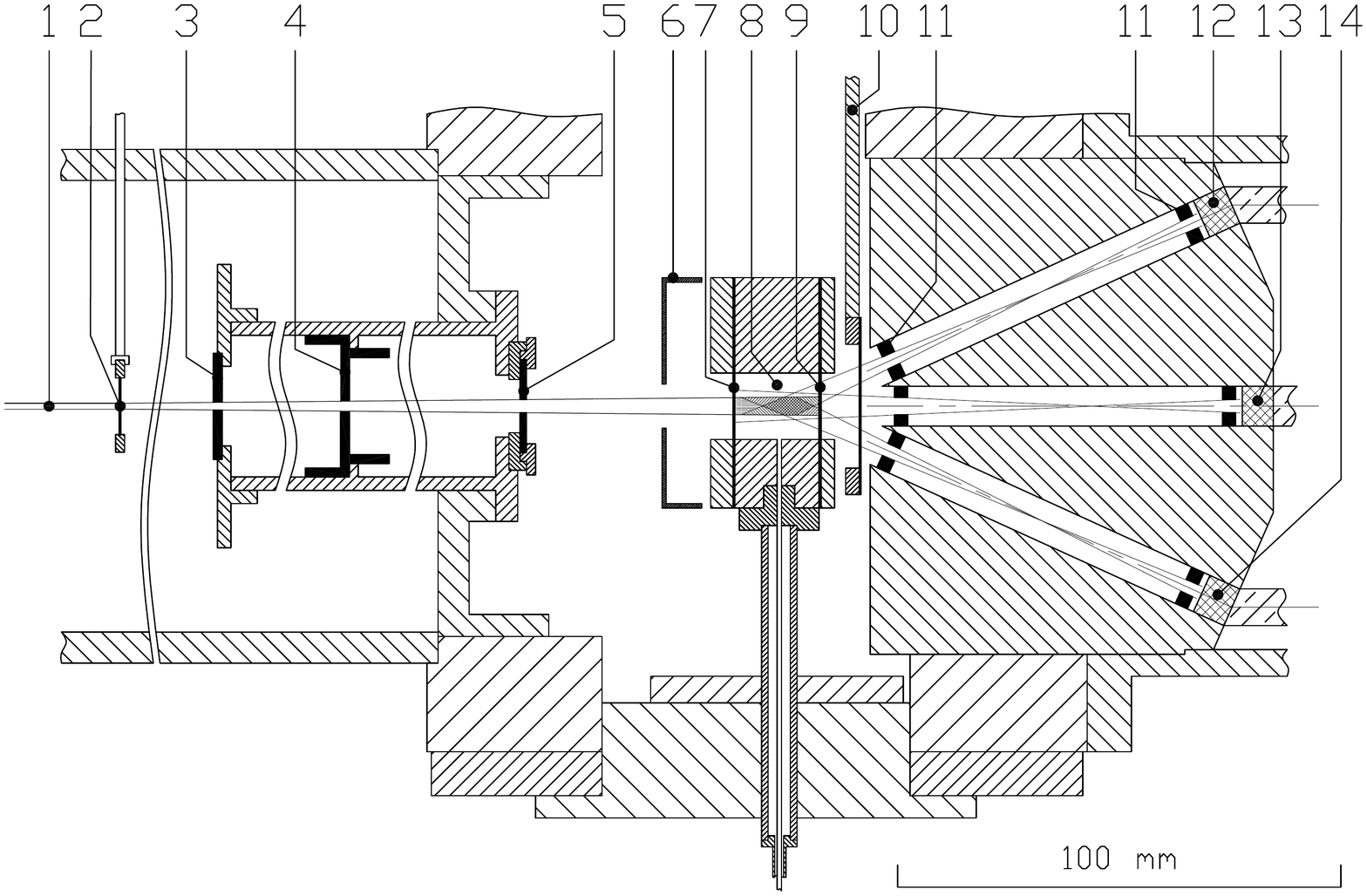}
\caption{View from the top along the horizontal midplane of the experimental setup: (1) unpolarized initial beam ({\O}=1.5\,mm); (2) target foil(s) in an aluminum frame, (3-5) diaphragms $D_{1}$ ({\O}=2.0\,mm),  $D_{2}$ ({\O}=2.5\,mm), and $D_{3}$ ({\O}=3.0\,mm) positioned 132, 187, and 251\,mm behind the target; (6) electron-backbending electrode; (7) 6.5\,$\mu$m Havar front window \cite{Havar} of the polarimeter cell positioned 299\,mm from the target; (8) polarimeter cell of 18\,mm length and 14\,mm diameter filled with $^{3}\rm{He}$ gas of 3 bar; (9) 100\,$\mu$m tantalum rear window; (10) additional 300 $\mu$m tantalum foil on a sliding ladder inserted to stop deuterons during beam tuning; (11) tantalum apertures;  (12-14) three of the five  NaI(Tl) detectors with light guides to the Philips XP1911 secondary-electron multiplier tubes.}
\label{Setup}
\end{center}
\end{figure}

As shown in Fig.\,\ref{Setup}, the setup\,\cite{Engels_1997} behind the target position consisted of a set of three diaphragms, the polarimeter cell, and five detectors to measure protons from the polarimeter reaction $\vec{d}+{^{3}\rm{He}}\to p+{^{4}\rm{He}}$. The diaphragms confined the transmitted deuteron beam to the reaction cell, widened by multiple small-angle Coulomb scattering in the target. They were mounted electrically insulated and were used to monitor the current by deuterons and charged particles from deuteron-induced reactions.  The polarimeter cell of 18\,mm length and 14\,mm diameter was closed by a 6.5\,$\mu$m Havar front window \cite{Havar} and a 100\,$\mu$m tantalum rear window. It was filled with $\mathrm{^{3}He}$ gas of 3\,bar. The 100\,$\mu$m thickness of the tantalum rear window was sufficient to stop the residual deuterons, while the protons from the reaction were only slightly affected due to the high 18.3\,MeV $Q$ valueof the polarimeter reaction. The additional, 300\,$\mu$m thick tantalum foil could be inserted to stop the deuterons during tuning of the initial beam without target. The target-gas chamber with the 100\,$\mu$m tantalum foil and the additional 300\,$\mu$m tantalum foil, mounted electrically insulated from the other components, were used 
to monitor the beam current behind the diaphragms. The electrode in front of the cell, kept at a negative potential against the cell, was installed to reflect secondary electrons emitted from the Havar window. The initial beam for each target was tuned to deliver a current of about 7\,nA to the polarimeter cell. The current to the set of diaphragms then had values around 200\,nA.

The polarimeter was equipped with a  forward detector (labeled F) and four side detectors positioned at polar angles of $\theta$=24.5\,$^{\circ}$ and azimuthal angles $\varphi$=0\,$^{\circ}$ [labeled L (left)], 90\,$^{\circ}$ [U (up)], 180\,$^{\circ}$ [R (right)], and 270\,$^{\circ}$ [D (down)] as seen in beam direction. Their acceptances were defined by pairs of 3 mm thick tantalum apertures. All five front apertures were positioned 25\,mm from the intersection point of the central trajectories of the side detectors, located 13\,mm behind the Havar front window. The rectangular front-aperture slits of the side detectors were 3\,mm wide in the direction of the reaction plane and 6\,mm perpendicular to it. The circular front aperture of the forward detector had a diameter of 4\,mm. All five rear apertures were circular with diameters of 3\,mm. They were positioned 75 mm behind the front apertures near to the detectors. The aperture geometry limits the polar acceptance angles of the side detectors to $(24.5\pm2.4)$\,$^{\circ}$ and that of the forward detector to $\le2.1$\,$^{\circ}$.

The axis of the forward-transmitted deuteron beam coincides with that of the initial beam and is the symmetry or quantization $z$ axis.  In this case, the full formula for the differential, spin-dependent cross-section\,\cite{Ohlsen_1972} of the polarimeter reaction with proton emission under a polar angle $\theta$ reduces to
\begin{equation}
\sigma(E_{\rm c},\theta)=\sigma_{0}(E_{\rm c},\theta)\cdot
               \Big[ 1+\frac{1}{2}\cdot p_{zz}(E_{\rm c})\cdot A_{zz}(E_{\rm c},\theta) \Big].
\label{Sigma}
\end{equation}
Here $E_{\rm c}$ is the deuteron energy at the reaction vertex in the gas of the polarimeter cell and $\sigma_{\rm 0}(E_{\rm c},\theta)$ is the unpolarized differential cross section. The term $A_{zz}(E_{\rm c},\theta)$ is a  component of the tensor analyzing power of the polarimeter reaction in Cartesian description. 

The number of protons, counted in one of the detectors during a run time $T$ with a deuteron current $j_{c}(t)$ to the polarimeter cell, is
\begin{equation} 
N'_{\rm i}(E_{\rm c},\theta_{\rm i})=\rho_{\rm He} \cdot l_{\rm i} \cdot \epsilon_{\rm i}
                                            \cdot \sigma(E_{\rm c},\theta_{\rm i}) \cdot \int_{0}^{T} j_{\rm c}(t)dt \cdot f_{\rm DT,i}
\label{Counts}
\end{equation}
with i=L, R, U, D, and F. Here $\rho_{\rm He}$ is the number density of the  ${^{3}{\rm He}}$ in the reaction cell, $l_{\rm i}$ is the length of the volume along the beam, from where protons can be emitted into the detector according to the cut by the apertures, $\epsilon_{\rm i}$ is the detector efficiency, and $\sigma(E_{\rm c},\theta_{\rm i})$ is the the reaction cross section of Eq.\,(\ref{Sigma}) for $\theta_{\rm i}=0\,^{\circ}$ and 24.5\,$^{\circ}$, respectively. The last term $f_{\rm DT,i}<1$ stands for the loss of counts due to dead time of the detector branch. To achieve comparable counts in the runs, the measuring time $T$ of a run was defined by the charge to the polarimeter cell. A run was stopped when the number of pulses from a current integrator reached a preset value. To determine the dead-time losses, pulses of 1.0 MHz were fed into the electronic branch of the five detectors. The dead-time factors $f_{\rm DT,i}$ were derived from the ratio of the pulses, stored by the data-acquisition system, to the total number pulses, fed during $T$. For all five detectors 1-$f_{\rm DT,i}$ was around 1\%.

The cross section $\sigma(24.5^{\circ})$ does not depend on the azimuthal angle position of the four side detectors. This allows one to combine the dead-time corrected counts $N_{\rm i}(E_{\rm c})=N'_{\rm i}(E_{\rm c},\theta_{\rm i})/f_{\rm DT,i}$ of the four side detectors (i=L, U, R, D and $\theta_{\rm i}=24.5\,^{\circ})$ and to formulate the ratio to the dead-time corrected counts in the forward detector (i=F and $\theta_{\rm F}=0\,^{\circ})$,
\begin{samepage}
\begin{eqnarray}
r(E_{\rm c}) &= &\frac{N_{\rm L}(E_{\rm c})+N_{\rm R}(E_{\rm c})+N_{\rm U}(E_{\rm c})+N_{\rm D}(E_{\rm c})}
                                 {N_{\rm F}(E_{\rm c})}
                                           =\frac{\sum_{\rm i=L,U,R,D}l_{\rm i}\cdot \epsilon_{\rm i}}{l_{\rm F}\cdot \epsilon_{\rm F}}\cdot
                                         \frac{\sigma(E_{\rm c},24.5\,^{\circ})}{\sigma(E_{\rm c},0\,^{\circ})}\nonumber\\
                            &  = &\frac{\sum_{\rm i=L,U,R,D}l_{\rm i}\cdot \epsilon_{\rm i}}{l_{\rm F}\cdot \epsilon_{\rm F}}\cdot
                                     \frac{\sigma_{0}(E_{\rm c},\,24.5\,^{\circ})\cdot [1+\frac{1}{2}
                                     \cdot p_{zz}(E_{\rm c})\cdot A_{zz}(E_{\rm c},\,24.5\,^{\circ})]}
                                            {\sigma_{0}(E_{\rm c}, 0\,^{\circ})   \cdot [1+\frac{1}{2}
                                             \cdot p_{zz}(E_{\rm c})\cdot A_{zz}(E_{\rm c},0\,^{\circ})]}~.
\label{C_Ratio}
\end{eqnarray}
\end{samepage}
By the summation count-rate fluctuations are compensated, which were observed in the single side detectors as up to $\sim5\%$ due to instabilities of the initial beam.

The relation of Eq.\,(\ref{C_Ratio}) would allow one to determine $p_{zz}(E_{\rm c})$ of the beam behind a carbon target from the measured ratios $r^{\rm C}(E_{\rm c})$ on the premises that, besides the unpolarized cross sections and analyzing powers of the polarimeter reaction, the values of $l_{\rm i}\cdot \epsilon_{\rm i}$ are known. Measurements, however, with unpolarized beams into the polarimeter cell ($p_{zz}\equiv 0$) yield reference ratios $r^{\rm ref}(E_{\rm c})$ determined by the products $l_{\rm i}\cdot\epsilon_{\rm i}$ and the unpolarized cross sections. The assumption is made that the ratio $\sum l_{\rm i}\cdot\epsilon_{\rm i}/l_{\rm F}\cdot\epsilon_{\rm F}$ has the same value in the measurements with the carbon targets and those to obtain the reference data. This means that the deuteron-trajectory distributions in the gas volume of the polarimeter cell are assumed to be the same in all measurements. The validity of the assumption, already used in our earlier analysis\,\cite{Seyfarth_2010}, is discussed in Sec.\,\ref{RefData}. Then a further ratio can be formulated,
\begin{equation}
\frac{r^{\rm C}(E_{\rm c})}{ r^{\rm ref}(E_{\rm c})}=
              \frac{1+\frac{1}{2}\cdot p_{zz}(E_{\rm c})\cdot A_{zz}(E_{\rm c},24.5\,^{\circ})}
                     {1+\frac{1}{2}\cdot p_{zz}(E_{\rm c})\cdot A_{zz}(E_{\rm c},0\,^{\circ})},
\label{Ratio}
\end{equation}
from which $p_{zz}(E_{\rm c})$ is obtained as
\begin{equation}
p_{zz}(E_{\rm c})
    =\frac{2\cdot \left[ \displaystyle 1-\frac{r^{\rm C}(E_{\rm c})}{r^{\rm ref}(E_{\rm c})} \right] }
   {\displaystyle \frac{r^{\rm C}(E_{\rm c})}{r^{\rm ref}(E_{\rm c})}\cdot A_{zz}(E_{\rm c}, 0\,^{\circ})-A_{zz}(E_{\rm c}, 24.5\,^{\circ})}.
\label{Pzz}
\end{equation}
With $R(E_{\rm c})=r^{\rm C}(E_{\rm c})/r^{\rm ref}(E_{\rm c})$ this relation yields $\Delta p_{zz}(E_{\rm c})$ caused by the variation of $\Delta R(E_{\rm c})$ as
\begin{equation}
\Delta p_{zz}(E_{\rm c})=\frac{-2}{R(E_{\rm c})\cdot A_{zz}(E_{\rm c},0\,^{\circ})-A_{zz}(E_{\rm c},24.5\,^{\circ})}
           \cdot \Big[ 1+\frac{A_{\rm zz}(E_{\rm c},0\,^{\circ})\cdot(1-R(E_{\rm c})}{R(E_{\rm c})\cdot A_{zz}(E_{\rm c},0\,^{\circ})
           -A_{zz}(E_{\rm c},24.5\,^{\circ})} \Big]\cdot \Delta R(E_{\rm c}).
\label{DeltaPzz}
\end{equation}
For the minimum and maximum energies $E_{\rm c}$=4.78 and 8.24\,MeV in the present measurements, the analyzing powers are $A_{zz}(0\,^{\circ})=-1.74$ and $-1.63$\,\cite{Schmelzbach_1976,Tonsfeldt_1980, Engels_1997}, respectively, and $A_{zz}(24.5\,^{\circ})=-0.81$ and $-0.98$\,\cite{Tonsfeldt_1980,Bittcher_1990,Engels_1997}, respectively. With the average values $\overline{A_{zz}(0\,^{\circ})}=- 1.69$, $\overline{A_{zz}(24.5\,^{\circ})}=-0.89$, and $\overline{R}=0.95$ (Fig.\,2 of\,\cite{Seyfarth_2010}), one obtains the approximate relation
\begin{equation} 
\Delta p_{zz}\sim 3.2\cdot \Delta R.
\label{dPzz/dR}
\end{equation}
It allows to estimate to variation of $p_{zz}$ due to uncertainties in the ratios $r^{\rm C}$ and $r^{\rm ref}$. The values of $p_{zz}$ lie between about $-0.3$ and 0, the ratio $R(E_{\rm c})=r^{\rm C}(E_{\rm c})/r^{\rm ref}(E_{\rm c})$ varies between 0.9 and 1.0. An uncertainty $\Delta R=\pm 0.01$ results in a significant $\Delta p_{zz}$ of about $\pm 0.03$. The uncertainties are discussed in Sec.\,\ref{Uncertainties}.
\subsection{Targets and beam energies~\label{Targets+Beams}}
For all measurements the energies of the initial deuteron beam were chosen such that the deuteron beam in the reaction cell had energies $E_{\rm c}$ between 5 and 8\,MeV. This choice is based on two facts.
\begin{enumerate}
\item The unpolarized differential cross section $\sigma_0(E_{\rm c},\,24.5\,^{\circ})$ decreases smoothly with increasing energy in the energy range 4.78 to 8.24\,MeV, whereas  $\sigma_{0}(E_{\rm c},\,0\,^{\circ})$ is almost constant\,\cite{Bittcher_1990,Engels_1997}. Thus, for an unpolarized beam into the polarimeter cell, i.e, $p_{zz}(E_{\rm c})=0$ in Eq.\,(\ref{C_Ratio}), the measured ratio $r^{\rm ref}(E_{\rm c})$ decreases smoothly with increase of $E_{\rm c}$.
\item Between $E_{\rm c}$=4.78 and 8.24\,MeV the tensor analyzing powers given above are large and vary smoothly. To be used in Eq.\,(\ref{Pzz}), the existing data for $A_{zz}(E_{\rm c},0\,^{\circ})$ and  $A_{zz}(E_{\rm c},24.5\,^{\circ})$ were fitted by polynomials.
\end{enumerate}

The K\"oln tandem accelerator provided unpolarized deuteron beams of energies up to $\sim$20\,MeV. After energy loss in the target and the Havar front window of the reaction cell, the beam energy $E_{\rm c}$ was requested to lie between about 5 and 8\,MeV. This restricted the areal graphite-target density to about $\le200$\,mg/cm$^2$.  Seven targets of different areal density were used, three in the runs in 2003 and four in 2006 (Table\,\ref{Targets}). The target foils of
\begin{table}
\caption{Target material M and areal target density $\delta$, target label, year of experiment, minimal and maximal initial deuteron beam energy $E_{\rm in}$, calculated mean beam energies $E_{\rm out}$ behind the target and $E_{\rm c}$ in the $^{3}{\rm He}$ polarimeter cell. The arrows indicate the direction in which $E_{\rm in}$ was changed in steps of 0.1\,MeV.} 
\begin{center}
\begin{tabular}{c c c c c c c c c c c c}\hline\hline
    &                                        &         &          & \multicolumn{3}{c}{$E_{\rm in}$(MeV)} &
&  \multicolumn{2}{c}{$E_{\rm out}$(MeV)}  &   \multicolumn{2}{c}{$E_{\rm c}$(MeV)}  \\
M  & $\delta$(mg/cm$^2$) & ~label~ &    &  min  &                      &  max                &
   &     min   &  max                  &             min   &max      \\\hline

no &                                       &    MT &  2003  & ~~6.00 & $\leftarrow$ & ~7.30       &   & \multicolumn{2}{c}{$\equiv E_{\rm in}$ }         &    5.54 & 6.90  \\

Au &    ~~$5.0\pm0.3$                &   Au5 & 2006  & ~~6.20 & $\leftarrow$ & ~7.90       &        &  6.02 & 7.74   &    5.56 & 7.36 \\

Au &   ~~$10.0\pm0.3$               &  Au10& 2003 &  ~~6.80 &  $\rightarrow$ & ~7.80       &        &  6.46 & 7.49  &    6.02 & 7.09\\

C  &  ~$35.90\pm0.19$              &   C36 & 2006 & ~~9.50 & $\leftarrow$ & 10.50      &        &   6.49 &7.79  &     6.06 & 7.41 \\

C  &  ~$57.69\pm0.32$              &   C58 &  2003  & ~10.80 & $\rightleftharpoons$ &~12.50\footnotemark[1]     &        &  6.18 & 8.60  &     5.73 & 8.24 \\
 
C  &  ~$93.59\pm0.37$              &   C94 & 2006  & ~13.00 & $\leftarrow$ &14.00       &        &  6.17 & 7.84   &    5.72 & 7.46 \\

C  & $129.49\pm0.42$             & C129 & 2006  & ~14.80 & $\leftarrow$ &15.90        &       &  5.88 & 7.93  &     5.41 & 7.55 \\

C  & $152.63\pm0.75$             & C153 &  2003  & ~15.70 & $\rightleftharpoons$ & 16.70\footnotemark[2]      &       &  5.28 & 7.37  &     4.78 & 6.97 \\

C  & $165.39\pm0.46$            & C165 &  2006& ~16.70 & $\leftarrow$ & 17.50       &      &  6.16 & 7.77  &     5.70 & 7.39 \\
 
C  & $187.93\pm0.74$            &C188~&  2003 & ~17.50 & $\rightarrow$ & 18.60      &    & 5.60 &7.97  &   5.12 &7.59~ \\\hline\hline
\end{tabular}
\footnotetext[1]{run sequence 12.50 to 12.10\,MeV,  11.00 to 12.00\,MeV, 10.90\,MeV, and finally 10.80\,MeV}
\footnotetext[2]{run sequence 15.70\,MeV, 16.50\,MeV, 16.30\,MeV followed by runs with 16.70 to 16.20\,MeV}
\end{center}
\label{Targets}
\end{table}
 $3\times3$\,cm$^2$ were cut from graphite sheets produced by rolling out expanded graphite\,\cite{ExpGra} to foils of nominal thickness  of 0.2, 0.3, and 0.5\,mm and nominal mass density of 0.7, 1.0, and 1.3\,g/cm$^3$, respectively\,\cite{NGS}. The areal densities of the targets and their uncertainties, caused by inhomogeneities in the graphite sheets, were determined by averaging over about 100 circular samples cut from positions distributed over each of the three $20\times 30$\, cm$^{2}$ sheets. The obtained distributions were fitted by Gaussian distributions. Except for C36 and C58, the areal densities of the graphite targets of Table\,\ref{Targets}, are those of target-foil stacks. Their uncertainties result from quadratic superposition of the standard deviations determined for the three graphite sheets. The table includes the measurements without target and with thin gold foils, performed to obtain data for the reference count-number ratio $r^{\rm ref}(E_{\rm c})$ in Eq.\,(\ref{Pzz}). The thin gold targets were utilized to simulate multiple small-angle Coulomb scattering in the carbon targets under the assumption that no polarization is produced in the transmitted beam. The validity of this assumption is discussed in Sec.\,\ref{RefData} by comparison of the data measured with the gold targets and without target.

As a standard procedure, the initial energy from the accelerator, $E_{\rm in}$, was varied in steps of 0.1\,MeV in the range between the minimum and maximum values given in Table\,\ref{Targets}. The mean deuteron energies behind the targets, $E_{\rm out}$, and those in the polarimeter gas 13\,mm behind the entrance window, $E_{\rm c }$, were calculated with the use of a program based on the formulae of\,\cite{Andersen+Ziegler}. The energy differences from those obtained with the computer code STOP\,\cite{STOP} are below 0.01\,MeV. The arrows in Table\,\ref{Targets} indicate the direction of the energy changes in the sequence of runs with each of the  targets. The run sequence is of importance in so far as a change of the areal target density during the sequence of runs, caused by the beam, would lead to deviations from the energies $E_{\rm c}$, calculated under the assumption of constant areal density. As a consequence, incorrect values of $p_{\rm zz}(E_{\rm c})$ would be deduced with the use of Eq.\,(\ref{Pzz}), where the calculated values of $E_{\rm c}$ determine the values of the analyzing powers and $r^{\rm ref}(E_{\rm c})$. This is discussed in Sec.\,\ref{TimeDependence}.
\subsection{Properties of the graphite foils~\label{Prop_Foils}}
The nominal 99\% carbon purity of the three graphite sheets, specified by the deliverer\,\cite{NGS}, was confirmed by own analyses employing standard techniques\,\cite{ZCH}. The mass fractions of the most abundant impurity elements oxygen and sulphur, measured for the 0.3\,mm thick sheet, are collected in Table\,\ref{AnalyseZCH}. Only the sum is given for the mass fraction of the less frequent elements fluorine to barium. With the use of the atomic weights and the natural isotope abundances the atom fractions were calculated, partitioned to isotopes of nuclear spin equal to zero and different from zero. Again, for fluorine to barium only the resulting sums are given. Similar atom fractions were measured for the 0.2 and 0.5\,mm thick sheets. The data of the table demonstrate that nuclei of spins different from zero, mainly due to the natural abundance of ${^{13}{\rm C}}$, only make up slightly more than 1\% in the targets used to study the deuteron interaction with the spin-zero ${^{12}{\rm C}}$ nuclei.
\begin{table}
\caption{Analysis results for the 0.3\,mm graphite sheet\,\cite{ZCH}.  The atom fractions are calculated from the measured mass fractions  regarding for the atomic weights. The last columns show the partitioning of the atom fractions according to their abundances to isotopes with nuclear spin equal to 0 and different from 0. For the two most frequent impurity elements, O and S, the errors of the measured mass fractions are less than 2\%. For the less frequent impurities (F, Na, Mg, Al, Si, Cl, K, Ca, Ti, Mn, Fe, Zr, and Ba) only the sums are given. The errors for a single element are up to 20\%.}
\begin{center}
\begin{tabular}{c c c c c c c c c}
\hline\hline
element &~~atomic~~& measured & atom   &~isotope~&  isotope & nuclear & \multicolumn{2}{c}{atom fraction [\%]}\\
             & weight &mass fraction &~fraction~ &      &~abundanc~ &  spin   &       \multicolumn{2}{c}{with nuclear spin}\\
             &           & [\%]             &     [\%]     &         &    [\%]    &         &    ~~ = 0~~  &   $\neq$ 0\\\hline
 C      & 12.011 & 99.047   & 99.443   & $^{12}$C & 98.893    &   0      &    98.342  &     \\
        &        &          &          & $^{13}$C &  1.107    &  1/2    &        & 1.101    \\
 O      & 15.999 &  0.557   &  0.418   & $^{16}$O & 99.519    &  0       &    0.416  &     \\
        &        &          &          & $^{17}$O &  0.075   &  5/2    &       & 3.1$\cdot10^{-4} $    \\
        &        &          &          & $^{18}$O &  0.407   &  0    &~1.7$\cdot10^{-3}~$    &     \\
S & 32.060 & 0.175 & 0.066 & $^{32}$S & 95.020 &  0  & 0.062  &     \\
  &        &       &       & $^{33}$S &  0.750 & 3/2 &        & ~ 4.9$\cdot10^{-4}$ ~  \\
  &        &       &       & $^{34}$S &  4.215 &  0  & 2.8$\cdot10^{-3}$    &     \\
  &        &       &       & $^{36}$S &  0.017 &  0  & 1.1$\cdot10^{-5}$ &    \\
 F .. Ba & & 0.220 & 0.073 &  &  &  & 0.042& 0.031    \\
\bf total  & & \bf 100 & \bf 100 &  &  &  & \bf 98.867 & \bf 1.133  \\
\hline
\end{tabular}
\end{center}
\label{AnalyseZCH}
\end{table}
A crystalline structure might influence the deuteron-target interaction. Graphite sheets, produced by rolling out expanded graphite, show a distinct anisotropy of thermal conductivity, electric resistivity, and thermal expansion\,\cite{SGL}.  These properties reflect those of crystalline graphite, where they exist due to the pronounced layer structure of the crystal. In the graphite sheets the anisotropy is caused by the mechanical pressure put on the expanded material during the roll-out procedure, leading to a partial orientation of the carbon layers. The amount of crystalline structure in the three graphite sheets has been investigated by measuring transmission X-ray diagrams\,\cite{ZCH}. The widths of the observed reflexes and the strength of the background were compared with values measured for graphitic carbons in the transition from the three-dimensional crystalline graphite to the non-graphitic carbons of random layer structure\,\cite{Franklin_1951}. Obviously, the sheets consist of comparable parts of graphite-like layers oriented parallel to the foil surface, disoriented layers, and an amorphous fraction. This finding is supported by the fact that the mass density of about 1\,g/cm$^3$ of the sheets is appreciably lower than that of crystalline graphite, which is 2.265\,g/cm$^3$~\cite{Howe_2003}. Accretion of the crystalline phase has been observed at temperatures above 1700\,$^{\circ}$\,C\,\cite{Franklin_1951} only. According to estimations, however, the temperature in the target beam spot due to energy loss of the deuterons never exceeded $200\,^{\circ}$C. Furthermore, all graphite targets except C36 and C58 consisted of stacks of foils fixed in the target frames without special care of parallel adjustment. All this allows the conclusion that coherent solid-state effects in the deuteron-target interaction due to a crystalline target structure can be excluded.
\section{Measurements and results~\label{Meas+Res}}
\begin{figure}
\begin{center}
\includegraphics[width=8cm]{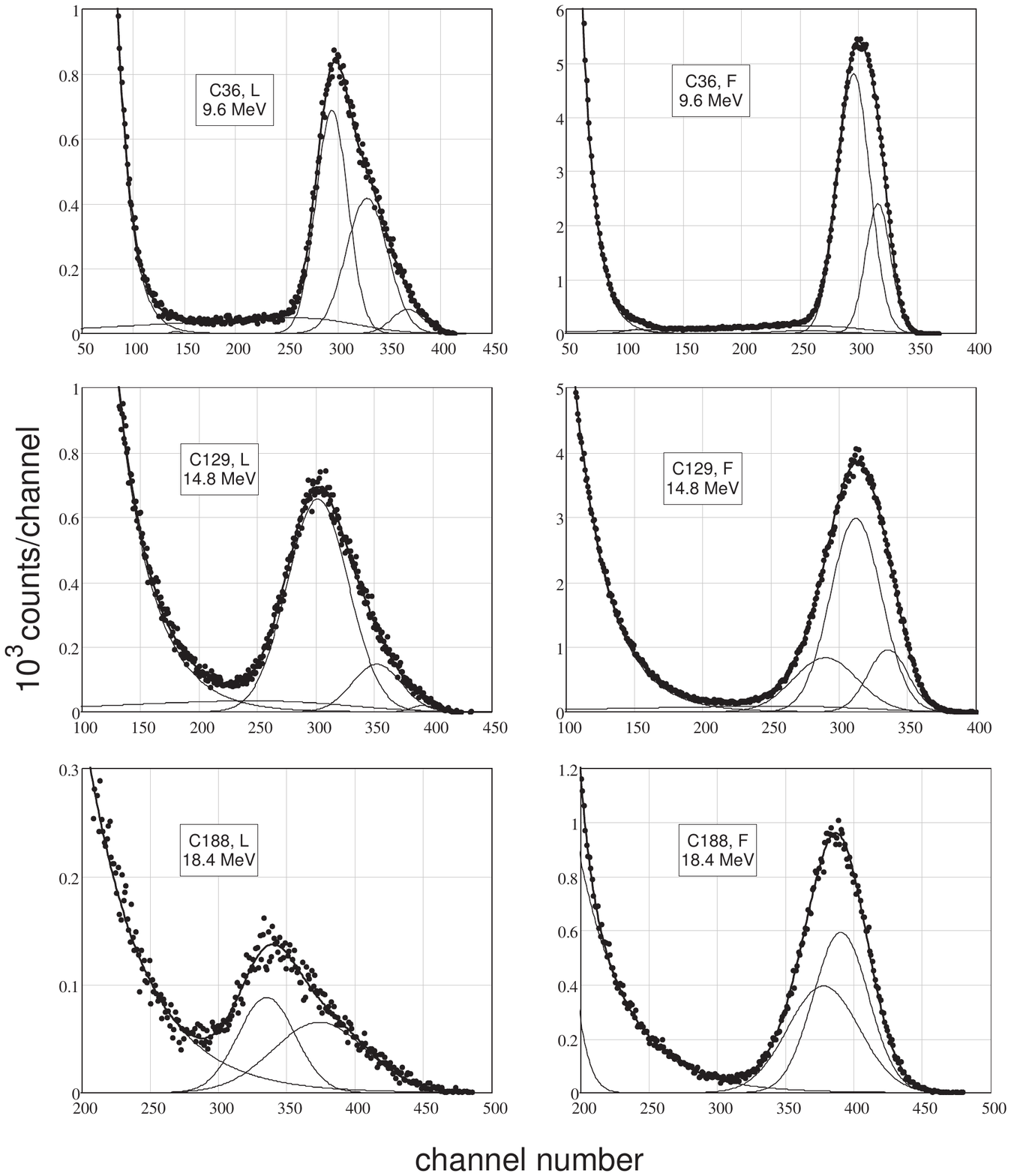}
\caption{Examples of proton spectra from the polarimeter reaction $\vec{d}+^3$He$\rightarrow$p+$^4$He, measured with the left side (L) and the forward (F) detector, together with the results of the fits (heavy lines) and the components of the fit functions (thin lines). These are three Gaussians to account for asymmetries of the peaks together with an exponential and a modified error function for the background. The inserts give the targets and the initial deuteron-beam energies.}
\label{Spectra}
\end{center}
\end{figure}

Six characteristic proton spectra are collated in Fig.\,\ref{Spectra}. The peaks, caused by protons from the polarimeter reaction $\vec{d}+^3$He$\rightarrow$p+$^4$He, were fitted by up to three Gaussian distributions to account for asymmetries. The low-energy background, originating from electronic noise and inelastic processes in the target and the polarimeter components, was fitted by an exponential function. The flat background component stems from protons scattered off the collimators into the detector and from incomplete light collection 
in the detector system. This component was fitted by a modified error function under the peak with a linear decrease towards zero energy. In the spectra, taken for the reference data and with the targets C36 to C129, the exponential background and the flat background could be  separated in the fit. With increasing target thickness, the exponential background under the proton peak is increasing. For the three thickest targets C153, C165, and C188 the flat background could not be resolved from the exponential component as is seen in the two lower spectra. The jeopardy by systematic errors in the fits is discussed in Sec.\,\ref{FitErrors}.
\subsection{Reference data~\label{RefData}}
The dependence of $p_{\rm zz}$ on the ratio  $r^{\rm C}(E_{\rm c})/r^{\rm ref}(E_{\rm c})$ as given in Eq.\,(\ref{Pzz}) is valid under the assumption that  the vertex distribution in the gas cell of the polarimeter is similar in the measurements with the carbon targets and those to achieve the reference data. The measurements with the gold targets Au5 and Au10 were performed to simulate the multiple Coulomb small-angle scattering by the carbon targets under the assumption that no polarization is produced.The validity of this assumption can be studied by comparing the measured ratios  $r^{\rm Au5}(E_{\rm c})$ and $r^{\rm Au10}(E_{\rm c})$ with $r^{\rm MT}(E_{\rm c})$ measured without target. Due to a slight deviation from a linear dependence on $E_{\rm c}$, the common data set is fitted by a second-order polynomial
\begin{figure}
\begin{center}
\includegraphics[width=8cm]{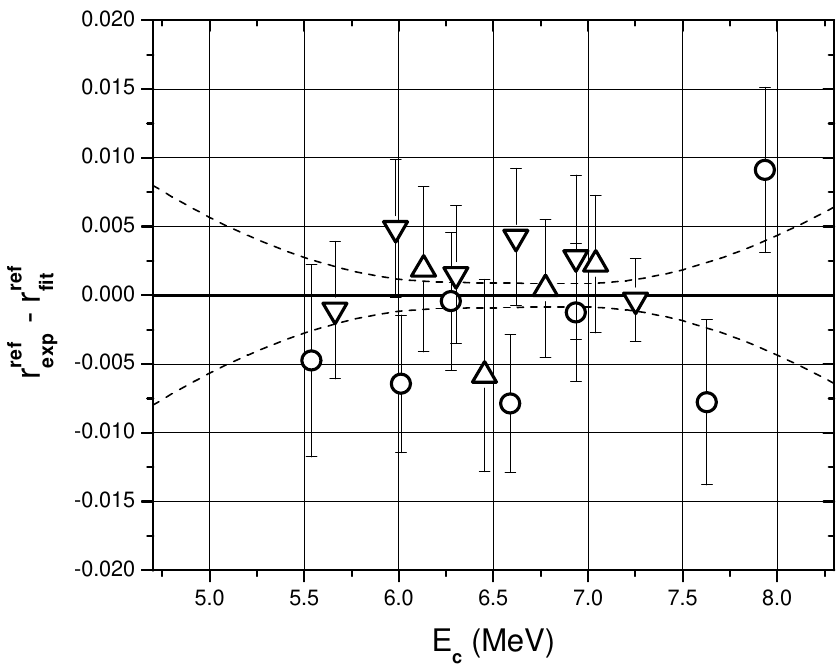}
\caption{Residuals in a second-order polynomial fit to the combined set of $r^{\rm MT}(E_{\rm c})$ ($\bigcirc$), $r^{\rm Au5}(E_{\rm c})$ ($\bigtriangledown$), and $r^{\rm Au10}(E_{\rm c})$ ($\bigtriangleup$) with the $\pm 1\sigma$ error band. To ease the readability data points are combined.}
\label{Res_Quadfit}
\end{center}
\end{figure}
contrary to the first analysis\,\cite{Seyfarth_2010}, where a linear fit had been applied to the Au5 ratios only. The residuals and the $\pm 1\sigma$ band are shown in Fig.\,\ref{Res_Quadfit}. With an average residual of $(+0.8\pm 0.7)*10^{-3}$, the $r^{\rm Au5,Au10}(E_{\rm c})$ lie higher than $(-3.5\pm 1.4)*10^{-3}$ for the $r^{\rm MT}(E_{\rm c})$. This can be interpreted as polarization production by the gold foils. Application of Eq.\,(\ref{Pzz}) with a second-order polynomial fit to the ratios $r^{\rm MT}(E_{\rm c})$ yields an $E_{\rm c}$-independent average value $p_{zz}^{\rm Au5,Au10}=(0.015\pm 0.014)$.

The difference can also be explained by a wider trajectory distribution in the polarimeter gas cell with the gold targets, which would lead to a higher acceptance by the side detectors and an increase of $r^{\rm Au5}(E_{\rm c})$ and $r^{\rm Au10}(E_{\rm c})$ against $r^{\rm MT}(E_{\rm c})$. Without target, the beam is not affected by the diaphragms. Here, the angular distribution of the trajectories in the polarimeter cell results from multiple small-angle Coulomb scattering in the 6.5\,$\mu$m Havar entrance window (5.4\,mg/cm$^2$, average mass number $\bar{\rm A}$=61,  average charge number $\bar{\rm Z}$=28\,\cite{Havar}). The angular width (FWHM) is calculated with the use of the formulae of\,\cite{PPB_2014} as 3.0\,$^{\circ}$ for $E_{\rm in}$=6.0\,MeV and 2.2\,$^{\circ}$ for $E_{\rm in}$=8.3\,MeV. Comparable widths are calculated for the widths behind the two gold targets. The trajectory angles against the beam axis, however, are cut by the diaphragms between target and polarimeter cell to $<0.51\,^{\circ}$. Therefore, with the gold foils the trajectory distributions in the polarimeter cell behind the Havar foil should be similar to that without target. The same holds for the carbon targets, where the widths of the trajectory distributions behind the targets are between 2.6\,$^{\circ}$ with C36 and 3.9\,$^{\circ}$ with C188. The angular cut by the diaphragms and the scattering by the Havar window justify the assumption that the ratio $\sum l_{\rm i}\cdot\epsilon_{\rm i}/l_{\rm F}\cdot\epsilon_{\rm F}$ has the same value in the reference and the carbon measurements, which leads to the Eqs.\,(\ref{Ratio}) and (\ref{Pzz}). The fit function $r^{\rm ref}_{\rm fit}(E_{\rm c})={\rm a}+{\rm b}\cdot E_{\rm c}+{\rm c}\cdot E_{\rm c}^2$ with a=1.7509, b=$-1.7535\cdot 10^{-1}$, c=4.5783$\cdot 10^{-3}$ ($E_{\rm c}$ in MeV) is used in Eq.\,(\ref{Pzz}) to determine the $p_{zz}(E_{\rm c})$. The errors according to the $\pm 1\sigma$ band enter the determination of $\Delta p_{zz}$.
\subsection{Carbon data and tensor polarization $p_{zz}$~\label{Carbon}}
\begin{figure}[b]
\begin{center}
\includegraphics[width=8cm]{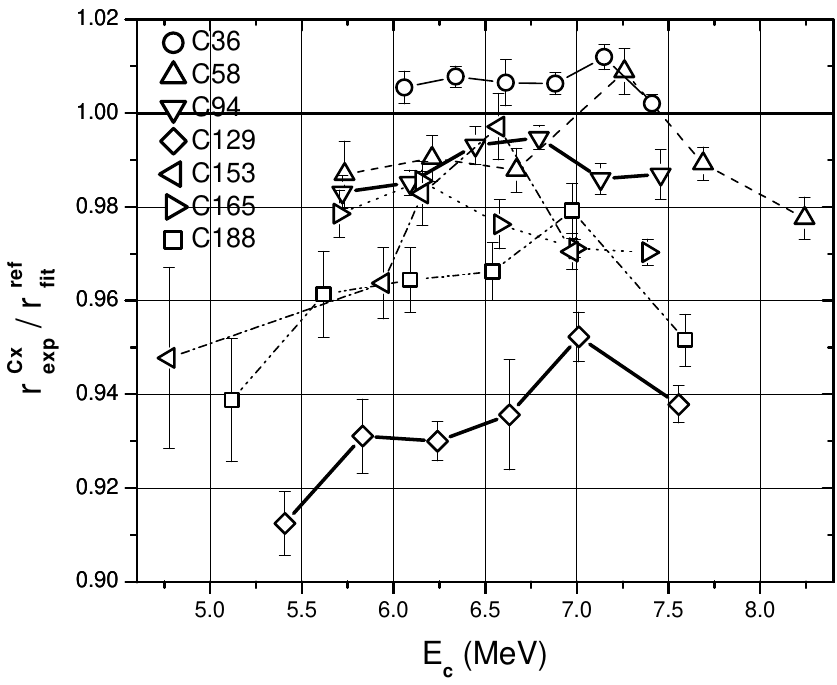}
\caption{Seven sets of ratios $r^{\rm Cx}_{\rm exp}(E_{\rm c})/r^{\rm ref}_{\rm fit}(E_{\rm c})$ resulting with the reference fit function $r^{\rm ref}_{\rm fit}(E_{\rm c})$ discussed in the preceding section. To ease the readability, only part of the measured ratios is shown covering the full range for each target. The error bars do not include those of the reference fit function.}
\label{R}
\end{center}
\end{figure}

With the seven carbon targets, the count ratios $r_{\rm exp}^{\rm Cx}(E_{\rm c})$ were measured with initial beam energies $E_{\rm in}$ spacing by 0.1\,MeV in the ranges given in Table\,\ref{Targets}. As the single exception, $E_{\rm in}$=11.7\,MeV is missing in the runs with the C58 target. At least two subsequent runs were performed with each energy to verify the beam stability. Figure\,\ref{R} shows ratios $r^{\rm Cx}_{\rm exp}(E_{\rm c})/r_{\rm fit}^{\rm ref}(E_{\rm c})$. For the targets C36, C58, and C94 the ratios lie around or near to one. A distinct difference between the ratios for C94 and those for C129 (strong full lines) is obvious. Those for the thicker 
targets C153, C165, and C188 lie between these cases. The estimate of Eq.\,(\ref{dPzz/dR}) can be used to interpret the data of this figure  concerning the expected $p_{zz}$ behind the targets. The difference of the average values $\overline{R(\rm C94)}-\overline{R(\rm C36)}$ of about $-0.02$ yields a change of $p_{zz}$ by $-0.06$ followed by a more distinct change of $-0.18$ due to the difference $\overline{R(\rm C129)}-\overline{R(\rm C94)}$ of about $-0.06$. Equation\,(\ref{dPzz/dR}) also shows the sensitivity of $p_{zz}$ on errors in the ratios $R$.

\begin{figure}
\begin{center}
\includegraphics[width=8cm]{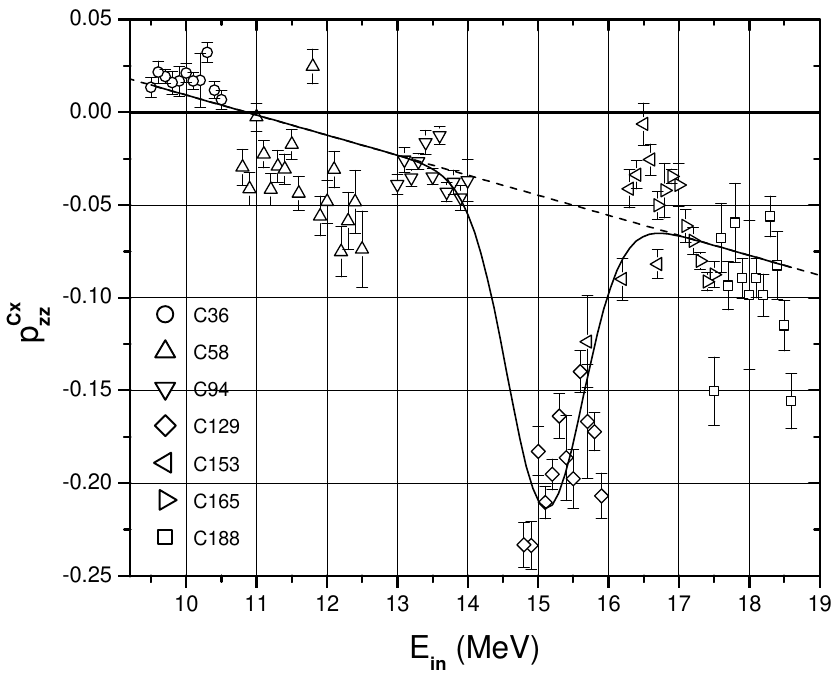}
\caption{The seven sets of tensor polarization $p_{\rm zz}^{\rm Cx}$ as a function of the initial beam energy $E_{\rm in}$ and their fit by a linear and a superposed Gaussian function. Repeated measurements with identical $E_{\rm in}$ are combined. The errors include those of the ratios $r^{\rm Cx}_{\rm exp}(E_{\rm c})$, those of the fit function $r^{\rm ref}_{fit}(E_{\rm c})$, and those of the analyzing powers. Unfortunately, no data exist in the gaps 12.5 to 13.0,  14.0 to 14.8, and 15.9 to 16.2\,MeV.}
\label{Pzz1}
\end{center}
\end{figure}

The set of the ratios $r^{\rm Cx}_{\rm exp}(E_{\rm c})/r^{\rm ref}_{\rm fit}(E_{\rm c})$ for the seven carbon targets, represented in Fig.\,\ref{R} by selected examples, with the use of Eq.\,(\ref{Pzz}) yields the the tensor polarizations $p_{zz}^{\rm Cx}(E_{\rm c})$ of the forward-transmitted beam. The special choice of the initial beam energies $E_{\rm in}$ for the seven targets to achieve similar beam energies $E_{\rm out}$ behind the targets allows to combine the $p_{zz}^{\rm Cx}$ in a common plot as a function of $E_{\rm in}$. This is done in Fig.\,\ref{Pzz1}. It shows the distinct dependence on the initial beam energy already presented in the corresponding figure of our first paper\,\cite{Seyfarth_2010}. The smooth energy  dependence there resulted from the fits to the reference data and, unnecessarily, to the seven carbon data sets, i.e., from the use of $r^{\rm Cx}_{\rm fit}(E_{\rm c})/r^{\rm ref}_{\rm fit}(E_{\rm c})$ in Eq.\,(\ref{Pzz}). In the present figure, the fluctuations in the carbon data are maintained by the use of the experimental values $r^{\rm Cx}_{\rm exp}(E_{\rm c})$. Due to the similarity of the beam energies behind the targets, the $p_{zz}$ as a function of $E_{\rm in}$ can be understood as a function of the areal carbon-target density. The linear fit function with a negative slope 
and the superposed Gaussian distribution, centered at 15.10\,MeV with a width (FWHM) of 1.3\,MeV, as well as the fluctuations clearly contradict the theoretical predictions of a weak and smooth dependence on the areal density with a single slope change around 11\,MeV.
\section{Uncertainties in the results~\label{Uncertainties}}
Systematic errors in the results of Fig.\,\ref{Pzz1} might be caused by
\begin{enumerate}
\renewcommand{\labelenumi}{\Alph{enumi}.}
\item uncertainties of the areal target densities given in Table\,\ref{Targets},
\item irradiation time dependent change of the areal target density in the beam-spot area,
\item errors in the fits to the proton spectra of the five detectors,
\item the variation of $E_{\rm out}$ behind the targets due to the change of the initial beam energy $E_{\rm in}$, and
\item energy-loss straggling in multiple small-angle Coulomb scattering.
\end{enumerate}
These possibilities are discussed in the following five subsections and summarized in subsection\,\ref{Conclusions_Uncertainties}
\subsection{Areal target densities~\label{NomDensity}}
The $p_{zz}$, given in Fig.\,\ref{Pzz1}as a function of $E_{\rm in}$, with the use of Eq.\,(\ref{Pzz}) are calculated from $r^{\rm C}_{\rm exp}(E_{\rm c})$, $r^{\rm ref}_{\rm fit}(E_{\rm c})$, $A_{\rm zz}(E_{\rm c},0\,^{\circ})$, and $A_{\rm zz}(E_{\rm c},24.5\,^{\circ})$. The $r^{\rm C}_{\rm exp}(E_{\rm c})$ are the ratios measured with the initial beam energy $E_{\rm in}$. The calculated energy losses in the targets, in the Havar window of the polarimeter gas cell, and in the $^3$He gas in the cell yield the values of $E_{\rm c}$, the mean deuteron energy in the polarimeter reaction. The values  from the fit function $r^{\rm ref}_{\rm fit}(E_{\rm c})$ and those of the analyzing powers are determined by the value of $E_{\rm c}$ calculated for the nominal values of the areal target densities $\delta_{\rm nom}$ of Table\,\ref{Targets}. In the beam-spot area, however, $\delta$ may be different from $\delta_{\rm nom}$. An increase (decrease) against $\delta_{\rm nom}$ would lead to a decrease (increase) of the real $E_{\rm c}$ against the value calculated for $\delta_{\rm nom}$ and the $E_{\rm c}$-dependence of $r^{\rm ref}_{\rm fit}(E_{\rm c})$, $A_{\rm zz}(E_{\rm c},0\,^{\circ})$, and $A_{\rm zz}(E_{\rm c},24.5\,^{\circ})$ would lead to a shift of the $p_{zz}$ to more negative (positive) values. The effect to the $p_{zz}^{\rm C129}$ is 
\begin{figure}
\begin{center}
\includegraphics[width=8cm]{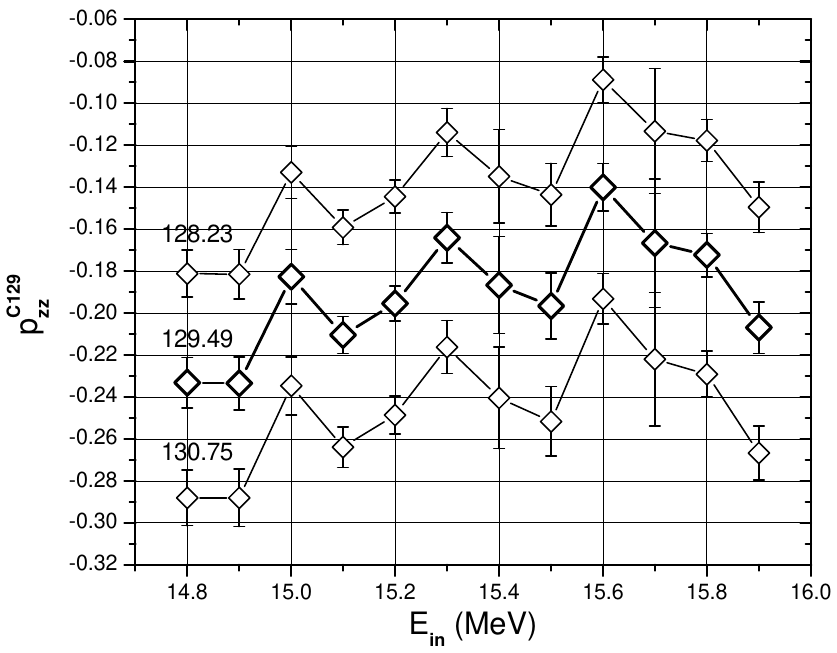}
\caption{The $p_{zz}^{\rm C129}$, obtained with the nominal areal C129 target density $\delta_{\rm nom}$= 129.49\,mg/cm$^{2}$ and those resulting under the assumption $\delta=\delta_{\rm nom}\pm 1.26$\,mg/cm$^{2}$, i.e., with increase and decrease of $\delta$ by three times the uncertainty given in Table\,\ref{Targets}. The run sequence was from $E_{\rm in}$=15.90 to 14.80\,MeV.}
\label{Pzz(C129)}
\end{center}
\end{figure}
demonstrated in Fig.\,\ref{Pzz(C129)}. The variation of $\delta$ against $\delta_{\rm nom}$ leads to overall shifts of $p_{zz}$. The fine structure,  however, is maintained. The same holds for the other targets.
\subsection{Time dependence of the areal target densities~\label{TimeDependence}}
Due to ionizing effects by the deuteron beam, the areal target densities $\delta$ might get modified during the sequence of runs with a target. According to the discussion of the preceding section, this would lead to errors in the deduced values of $p_{zz}$, increasing with the irradiation time. The necessity of corrections due to an increase of $\delta$ in $d-^{12}$C reactions ("carbon buildup") was mentioned earlier\,\cite{Baldeweg_1966}, \cite{Cords_1969_1}, and \cite{Jolivette_1974}, whereas no significant change was found in\,\cite{Ohlsen_1963}. As result of a systematic study\,\cite{Healy_1997}, an increase of $\delta$ is to be expected due to deposition of carbon atoms on the target surface confined to the irradiated area with a characteristic deposition rate of $1\cdot 10^{14}$\,C atoms\,cm$^{-2}$\,$\mu$Cb$^{-1}$.

Four carbon foils, used in the measurements in 2006 were still available. They show the characteristic brown color change in the beam spot area of 1.5\,mm diameter. In the run sequences with the four targets, the total charge $Q$ to the diaphragms behind the targets was about $3\cdot10^{7}$\,nCb. According to the above deposition rate\,\cite{Healy_1997} with the beam-spot area of 1.5\,mm diameter one expects a deposition of $5\cdot 10^{16}$\,C atoms or a tiny increase of the areal target density by about 10$^{-3}$\,mg/cm$^{2}$. To verify this prediction, circular pieces of 1.6\,mm diameter were stamped from the foils, in each case one encircling the beam spot and additional pieces from its neighborhood. The relative mass changes of the pieces with the beam spot against those from the neighborhood were determined as $(+2.8\pm$4.7)\% and $(+3.7\pm$4.8)\% for the two 36\,mg/cm$^{2}$ foils and $(+4.1\pm$4.8)\% and $(+1.0\pm$2.5)\% for the two 58\,mg/cm$^{2}$ foils. The results would allow a small increase of the areal target density. The correction,however, would not modify the fine structure of Fig.\,\ref{Pzz(C129)} and it therefore was neglected.
\subsection{Uncertainties in the fit results \label{FitErrors}}
\begin{figure}
\begin{center}
\includegraphics[width=8cm]{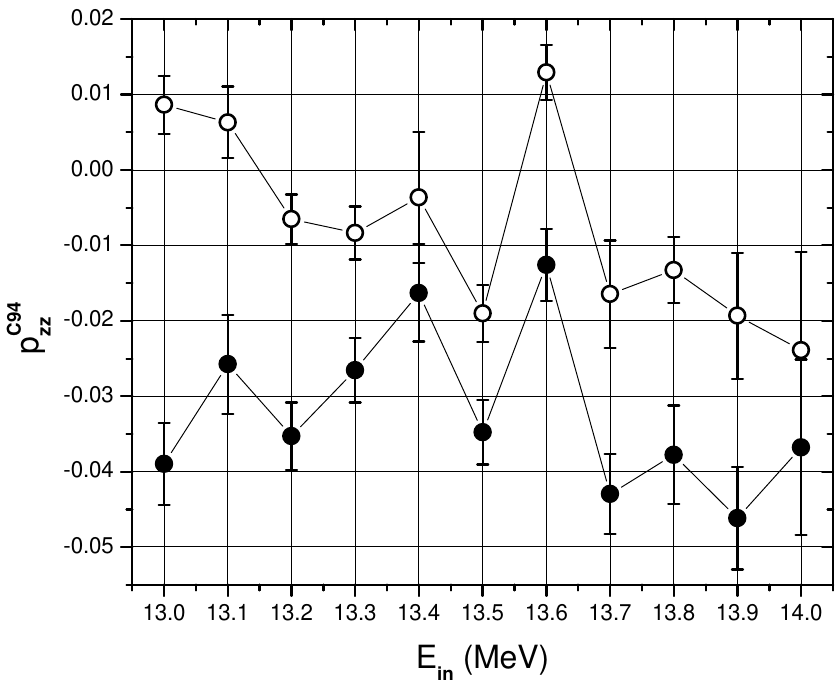}
\caption{The $p^{\rm C94}_{zz}$ resulting from the Gaussians, fitted to the peaks in the proton spectra (full circles), and those obtained from the sum of the Gaussians and the modified error function in the range of the Gaussians (open circles).}
\label{Pzz(C94)}
\end{center}
\end{figure}

With the carbon targets, the gold foils, and without target at least two runs were performed at the same $E_{\rm in}$ to account for instabilities of the initial beam. For each run, the proton peaks in the spectra measured with the four side detectors and the forward detector were fitted by up to three Gaussian functions to account for asymmetries in the peak shape. As background components the fits included an exponential function and a  modified error function (MEF) under the proton peak with a linear decrease towards zero energy. These account for inelastic processes in the detector surrounding and incomplete light collection in the detectors and protons scattered into the detectors by the tantalum apertures, respectively. As Fig.\,\ref{Spectra} shows, the separation of the exponential background and the proton peak gets worse with increasing areal target density. The overlap leads to increasing ambiguities in the fit results. For the three targets C153, C165, and C188 the exponential background dominates and the fits yield no contribution by the MEF. In the reference spectra with the gold target and without target, the proton peak is well separated and no ambiguities in the fits are encountered. Earlier experiments \cite{Reck_1989,Swillus_1990} have shown that the counts in the MEF carry the same information on the beam polarization as those in the peaks. After subtraction of the exponential background, the total counts in the energy range of the peaks should yield the same $p_{zz}$ as the counts in the peaks only, fitted by the Gaussian functions. Both sets of $p_{zz}$ for the C94 target are compared in Fig.\,\ref{Pzz(C94)}. The absolute values, resulting from the total counts in the peak ranges 
are systematically reduced against those from the peak counts only. Obviously the total counts contain an unpolarized component caused by incomplete subtraction of the unpolarized exponential background. The peak structure, however, is maintained. For the three targets C153, C165, and C188, where the fits yield the exponential background function only, the counts in the peak range may contain MEF components. This would the values of the $p_{zz}$ in a similar way.
\subsection{Variation of $E_{\rm out}$ with $E_{\rm in}$~\label{E_out}}
\begin{figure}
\begin{center}
\includegraphics[width=8cm]{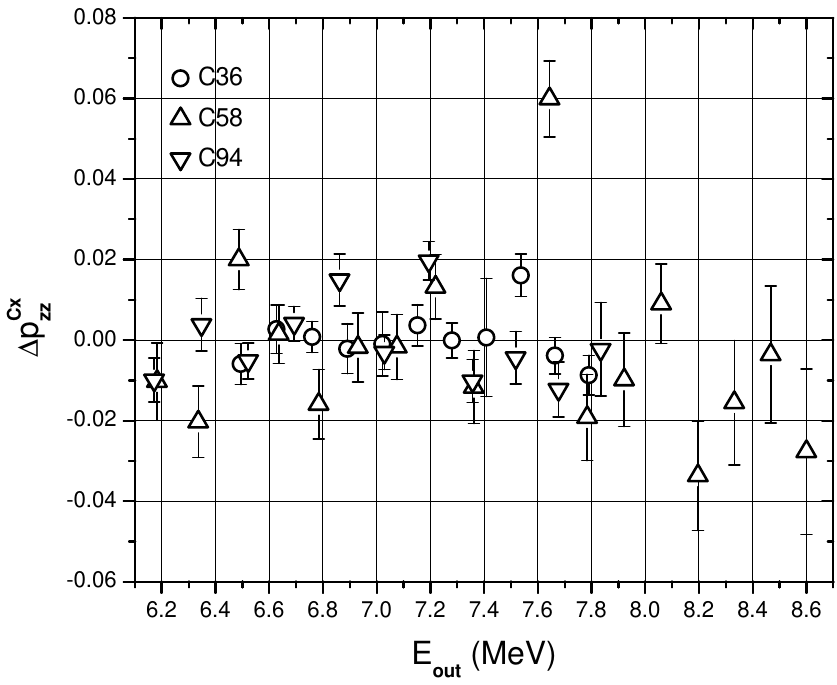}
\caption{Residuals of linear fits to the $p_{\rm zz}$ of Fig.\,\ref{Pzz1} for C36, C58, and C94 as a function of the energy $E_{\rm out}$ behind the targets.}
\label{Structure}
\end{center}
\end{figure}

It is an essential step in the present work to attribute the stepwise changes of $p_{zz}$ behind the targets to the $d-^{12}$C interaction in the energy intervals $\Delta E_{\rm in}$. Plotting $p_{zz}$ as a function of $E_{\rm in}$ as it is done in Fig.\,\ref{Pzz1} neglects two effects,
\begin{itemize}
\item the variation of the energy $E_{\rm out}$ with $E_{\rm in}$ in the runs with one of the targets and
\item the difference of $E_{\rm out}$ for one of the targets and the next thicker at the same value of $E_{\rm in}$.
\end{itemize}
As discussed in Sec.\,\ref{Targets+Beams}, the initial beam energies were chosen such that for all targets the $E_{\rm c}$ in the polarimeter cell come to lie between about 5 and 8\,MeV. The range for the mean beam energies behind the targets lies by about 0.5\,MeV higher between about 5.5 and 8.5\,MeV (Table\,\ref{Targets}). 
Significant contributions to the production of $p_{zz}$ in this energy range would show up as common structure obtained with different targets. For the comparison, Fig.\,\ref{Structure} shows the residuals of fits to the $p_{zz}$ of Fig.\,\ref{Pzz1} as a function of $E_{\rm out}$ for C36, C58, and C94. No counterpart, e.g., of the peak at
$E_{\rm out}$=7.64\,MeV for C58 is found for C36 and C94. This energy corresponds to the distinct peak at $E_{\rm in}$=11.80\,MeV observed with the C58 target (Fig.\,\ref{Pzz1}). In general, no significant common structure is visible. This justifies it to assign the changes of $p_{zz}$ to those of $E_{\rm in}$ for each target.

For the transition from one target to the next thicker one the effect due to the difference in $(E_{\rm out})$ can be studied by the $p_{zz}$ obtained with both targets C129 and C153 at $p_{zz}$=15.70\,MeV. With C129 $E_{\rm out}$=7.58\,MeV is appreciably higher than 5.28\,MeV with C153. The values $p_{zz}=-(0.140\pm 0.009)$ for C129 and $-(0.124\pm 0.025)$ for C153, however, agree within the errors. This justifies the combination of the $p_{zz}$, measured behind the C129 and the C153 targets, in a common plot as a function of $E_{\rm in}$.

At two other initial energies $p_{zz}$ was measured with adjacent targets. At $E_{\rm in}$=16.70\,MeV $p_{zz}=-(0.082\pm 0.008)$ with C153 and $p_{zz}=-(0.050\pm 0.007)$ with C165. At $E_{\rm in}$=17.50\,MeV $p_{zz}=-(0.150\pm 0.019)$ with C188 and $p_{zz}=-(0.088\pm 0.007)$ with C165. In both cases $p_{zz}$ with C165 is less negative. Maintaining the data for the C129, C153, and C188 targets, for both energies agreement is achieved by increasing of the C165 areal target density from the nominal value 165.39 to 166.32\,mg/cm$^2$ by two times the uncertainty 0.46\,mg/cm$^{2}$ of Table\,\ref{Targets}. For C165 this yields at $p_{zz}=-(0.087\pm 0.008)$ at $E_{\rm in}$=16.70\,MeV and $p_{zz}=-(0.129\pm 0.007)$ at $E_{\rm in}$=17.50\,MeV. Both values agree within the errors with those measured with C153 and C188.   The whole set of $p_{zz}$ measured with the C165 target is shifted correspondingly.

The weak polarizing effect in the energy range of $E_{\rm out}$ is confirmed by earlier works.  Polarized deuteron beams were slowed down from up to 7.5\,MeV to energies below 0.8\,MeV in polyethylene foils\,\cite{Seiler_1964} or from up to 6.8\,MeV to 0.4\,MeV in mylar foils\,\cite{Cords_1969_2} without measurable change of polarization. Thus, it is regarded as justified to neglect the polarization modification due to the variation of $E_{\rm out}$ and to combine the whole set of $p_{zz}$, measured with the seven targets, in a common plot versus the energy of the initial, unpolarized beam. 
\subsection{Energy-loss straggling~\label{Energy straggling}}
The effect of energy straggling produced in the targets and in Havar entrance window is discussed for the extreme value $p_{zz}=-(0.233\pm 0.012)$, measured with the C129 target at $E_{\rm in}$=14.80\,MeV and $E_{\rm c}$=5.407\,MeV. The width $\Gamma $of the (Gaussian) energy distribution around $E_{\rm c}$ in the polarimeter cell is calculated as 0.220\,MeV with the use of the formulae given in Ref.\,\cite{Segre_1964}. For $E_{\rm c}$ increased and decreased by $\Gamma/2$ one obtains $p_{\rm zz}(E_{\rm c}+\Gamma/2)=-0.193$ and $p_{\rm zz}(E_{\rm c}-\Gamma/2)=-0.275$, respectively. With equal weight (symmetric Gaussian distribution) the mean value $a\cdot p_{\rm zz}(E_{\rm c}+\Gamma/2)+b\cdot p_{\rm zz}(E_{\rm c}-\Gamma/2)$ with $a=b=1$ is $-0.234$. For assumed asymmetries $a=0.8, b=1.2$ and $a=1.2, b=0.8$ the mean values are $-0.242$ and $-0.226$, respectively. Both values agree with the measured $p_{zz}(E_{\rm c})=-(0.233\pm 0.012)$ within the error. The differences are caused by slight deviations of $r^{\rm ref}(E_{\rm c})$, $A_{zz}(E_{\rm c},0\,^{\circ})$, and $A_{zz}(E_{\rm c}, 24.5\,^{\circ})$ from linear dependence on $E_{\rm c}$. The asymmetry in the distribution of $E_{\rm c}$ would cause a weak, but continuous shift of the values of $p_{zz}$ as a function of $E_{\rm c}$ and also $E_{\rm in}$. The fine structure of $p_{zz}$ as a function of $E_{\rm in}$, however, would be maintained.
\subsection{Conclusions from Secs.\,\ref{NomDensity} to\,\ref{Energy straggling}~\label{Conclusions_Uncertainties}}
Due to the discussed uncertainties, the $p_{zz}$ of Fig.\,\ref{Pzz1} would be scaled up or down with a slight $E_{\rm in}$ dependence. The fluctuations of $p_{zz}(E_{\rm in})$ around the smooth energy dependences for the seven targets would be maintained in all regarded cases. The values of the $p_{zz}^{\rm C165}(E_{\rm in})$ of Fig.\,\ref{Pzz1} get adapted to the  $p_{zz}^{\rm C153}(E_{\rm in})$ and $p_{zz}^{\rm C188}(E_{\rm in})$ by an increase of the C165 areal target density by twice the uncertainty of Table\,\ref{Targets} or 0.6\% as discussed in Sec.\,\ref{E_out}. Adaptation of the $p_{zz}$, measured behind the C58 Target, to the C36 and C94 data by an arbitrary modification of the C58 areal target density is omitted. To adapt the C94 and C129 data would require for C94 to increase and/or for C129 to reduce the areal target density by more than five times the uncertainties given in Table\,\ref{Targets} to achieve the necessary values of $E_{\rm c}$. As a consequence, the areal target densities on the one hand of C36 and C58 and on the other hand of C153, C165, and C188 had to be modified. Therefore the adaptation of the $p_{zz}$ measured with the C94 and C129 targets would not make sense. 
\section{Modeling $p_{zz}(E_{\rm in})$ of the transmitted beam~\label{Modeling}}
\begin{figure}
\begin{center}
\includegraphics[width=8cm]{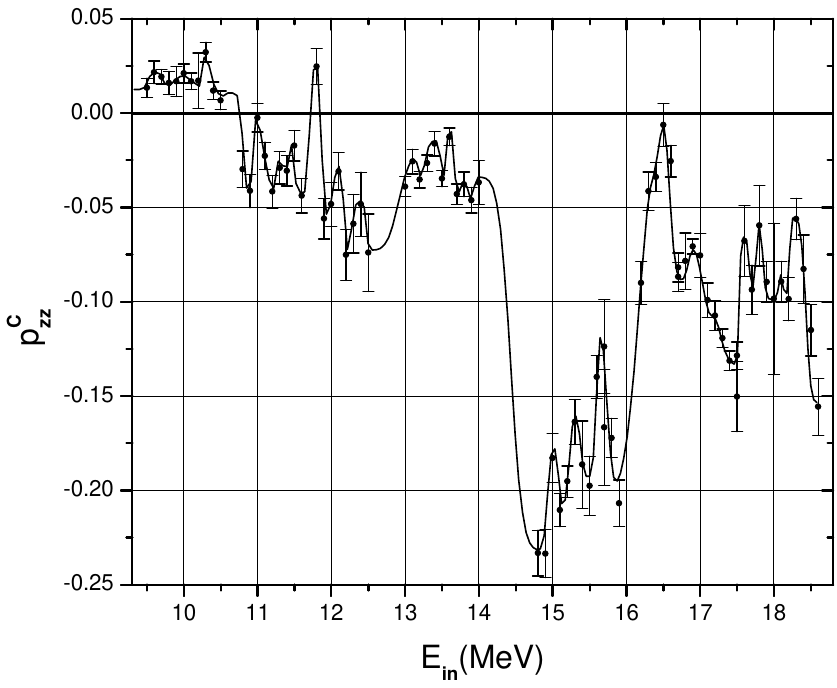}
\caption{Full points: $p_{zz}$ behind the carbon targets versus the initial energy of the deuteron beam . Full line: fit of the differential polarization production $\Delta p_{zz}/\Delta E_{\rm in}$ with the use of 51 Gaussian functions [Eq.\,(\ref{Gauss})] in Eq.\,(\ref{PzzSum2}), 25 with $\sigma_{\pm 1}=0, \sigma_{0}\ne 0$ to achieve $\Delta p_{zz}/\Delta E_{\rm in}>0$ and 26 with $\sigma_{\pm 1}\ne 0, \sigma_{0}=0$ for $\Delta p_{zz}/\Delta E_{\rm in}<0$. The fit parameters are collected in Tables\,\ref{Fitparameter-1} and \ref{Fitparameter-2}.}
\label{Pzz2}
\end{center}
\end{figure}

Figure\,\ref{Pzz2} again shows the $p_{zz}$ of Fig.\,\ref{Pzz1} measured behind the seven targets as a function of the initial beam energy $E_{\rm in}$. Compared to Fig.\,\ref{Pzz1}, the C165 data here are adapted to those measured with the C153 and C188 targets as founded in Sec.\,\ref{E_out}. As it is discussed in Sec.\,\ref{E_out}, 
the step-wise changes of $p_{zz}$ behind the targets can be associated with the polarization production $\Delta p_{zz}$ in the intervals $\Delta E_{\rm in}$. The step-wise increase of $E_{\rm in}$ by $\Delta E_{\rm in}$, however, is equivalent to an increase of the range of the deuteron energy $E$ in the target by $\Delta E=\Delta E_{\rm in}$. The change $\Delta p_{zz}$ is due to the $d-^{12}$C interaction in the energy interval $\Delta E$. Hence, $p_{zz}$=-0.156 measured behind the C188 target with $E_{rm in}$=18.60\,MeV (Fig.\,\ref{Pzz2}), can be understood as originating from the sequence of $\Delta p_{zz}(E)$ in the energy range between $E=E_{\rm in}$=18.60\,MeV and $E=E_{\rm out}$=9.50\,MeV.

When the integrals in Eq.\,(\ref{PzzThick2}) are replaced by sums over n$_{\rm out}$ thin layers of thickness $\Delta x$ (in units of length), the formula reads
\begin{eqnarray}
p_{zz}(\rho d)=\frac
  {2\cdot{\rm exp}\,\{-\rho \sum_{n=1}^{{\rm n_{out}}} \sum_{{\rm i}=1}^{{\rm N}_{\pm1}} \sigma_{\pm 1,{\rm i}}(E_{\rm n})\Delta x\}
  -2\cdot{\rm exp}\,\{-\rho \sum_{n=1}^{{\rm n_{out}}} \sum_{{\rm j}=1}^{{\rm N}_{0}}       \sigma_{0,{\rm j}}(E_{\rm n})\Delta x\}}
  {2\cdot{\rm exp}\,\{-\rho \sum_{n=1}^{{\rm n_{out}}} \sum_{{\rm i}=1}^{{\rm N}_{\pm1}} \sigma_{\pm 1,{\rm i}}(E_{\rm n})\Delta x\}
   +{\rm exp}\,\{-\rho \sum_{n=1}^{{\rm n_{out}}} \sum_{{\rm j}=1}^{{\rm N}_{0}}        \sigma_{0,{\rm j}}(E_{\rm n})\Delta x\}}.
\label{PzzSum2}
\end{eqnarray}
Here $\rho$ (in units of atoms per volume) is the number density of the carbon atoms in the targets. The $\sigma_{\pm 1,{\rm i}}(E_{\rm n})$ and $\sigma_{0,{\rm j}}(E_{\rm n})$ (in areal units) are the cross sections at the average energy $E_{\rm n}$ in the n$^{\rm th}$ layer, calculated  from $E_{\rm in}$ with the use of the Bethe-Bloch energy-loss formula\,\cite{Andersen+Ziegler}. The value of  $E_{1}$ lies near to $E_{\rm in}$, whereas $E_{\rm n_{out}}$ is near to $E_{\rm out}$, the beam energy behind the target. The analytical shapes of the N$_{\pm 1}$ cross sections $\sigma_{\pm1,i}(E_{\rm n})$ and the N$_{0}$ cross sections $\sigma_{0,j}(E_{\rm n})$ are chosen as Gaussian functions,
\begin{equation}
\sigma(E_{\rm n})=\sigma_{\rm tot}\cdot \frac{2.354}{\Gamma\sqrt{2\pi}}\,\cdot\,{\rm exp}\Big \{-\frac{1}{2}
      \Big [\frac{2.354(E_{\rm n}-E_{0}}{\Gamma}\Big ]^2\Big \}
\label{Gauss}
\end{equation}
with the normalization $\sum_{\rm n=1}^{\rm n_{out}} \sigma(E_{\rm n})\Delta E_{\rm n}(x)=\sigma_{\rm tot}$. Here the term $\Delta E_{\rm n}(x)$ regards for the change of the energy loss $\Delta E_{\rm n}$ in target-layers of constant thickness $\Delta x$. The numbers of Gaussian functions N$_{\pm 1}$ and N$_{0}$ depend on the achievement of a reasonable fit. The total cross sections $\sigma_{\rm tot}$ (in units of area$\cdot$energy), the full widths at half maximum $\Gamma$ and the central energies $E_{0}$ (both in units of energy) are the free fit parameters. They were varied "by hand" to yield a reasonable description of the data. The value of $p_{zz}$ at $E_{\rm in}$=9.50\,MeV was set to +0.0133 as measured with the C36 target.

The fit line of Fig.\,\ref{Pzz2} was achieved with the use of 26 cross-section combinations $\sigma_{\pm 1}(E)=0$, $\sigma_{0}(E)>0$ in the energy ranges with $\Delta p_{zz}(E)/\Delta E>0$ and 25 combinations $\sigma_{\pm 1}(E)>0$, $\sigma_{0}(E)=0$ for $\Delta p_{zz}(E)/\Delta E<0$. The fit parameters are collected in the columns one 
\begin{table}
\caption{Columns 1$-$3: parameters of the Gaussian functions [Eq.\,(\ref{Gauss})] in the fit of $p_{\rm zz}(E_{\rm in})$ of Fig.\,\ref{Pzz2} with the use of Eq.\,(\ref{PzzSum2}); column 4: tensor polarizations produced in the energy range of the single Gaussian function; column 5: peak energies $E_{\rm p}$ with errors in the $^{12}{\rm C}(d,\alpha_{2})^{10}{\rm B}$ excitation functions ($E_{0}$=9.52 to 13.93\,MeV\,\cite{Smith_1972}, 13.82 to 16.44\,MeV\,\cite{Jolivette_1974}, $E_{0}$=14.42, 16.14, 16.44, (16.65), 16.83\,MeV\,\cite{Ehrenstein_1971}); columns 6 and 7: weighted averages of $E_{0}$ and $E_{\rm p}$ and the corresponding excitation energies in $^{14}$N; columns 8$-$10: results of an S-matrix fit\,\cite{Jolivette_1981}.}
\begin{center}
\begin{tabular}{ccccccccccccc}\hline\hline
  1  &  2  &  3  &  4  &     &  5  &     &  6  &   7  &       &  8   &   9   &   10   \\
\multicolumn{4}{c}{present experiment}&~~~& $^{12}$C$(d,\alpha_{2})^{10}$B &~~~& & && \multicolumn{3}{c}{ $^{12}$C$(d,\alpha_{2})^{10}$B}      \\
$E_{0}$   & $\sigma_{\rm tot}$ & $\Gamma$ & $p_{\rm zz}$ & & $E_{\rm p}$ & & $\overline{E_{\rm d}}$ & $E(^{14}{\rm N}$*) &~~~~&$E_{\rm d}$& $E(^{14}{\rm N}$*)  & $J^{\pi}$ \\
  (MeV)    &   (b$\cdot$MeV)      &   (MeV)      &         & & (MeV)  && (MeV)  &   (MeV) &  & (MeV) &     \\\hline
    9.55     &   21  &  0.10  &     0.009      & &  9.523(0.025) & &  9.534(0.019)  & 18.433(0.016)  & &  9.53(0.07)   &   18.43(0.06)   &  $4^{+}$  \\
               &         &          &                   & &  9.625(0.007) &  &  9.625(0.007)    &   18.511(0.007)    & &   9.61(0.01)   &  18.50(0.01)   &  $5^{-}$   \\
               &         &          &                   &  & 9.640(0.017) &  &  9.640(0.017)    &   18.524(0.015)    & &   9.64(0.10)   &   18.52(0.08)  &  $2^{+}$  \\
               &         &          &           & &  9.665(0.015) &  &  9.665(0.015)    &   18.545(0.013)    & &   9.65(0.07)   &   18.53(0.06)  &  $3^{-}$  \\
    9.75  &  19  &  0.10  &  $-$0.009  & &  9.750(0.007) &  & 9.750(0.007)  & 18.618(0.006)  & &  9.77(0.08)   &   18.63(0.07)  &  $3^{-}$  \\   
    9.90     &   14  &  0.15  &     0.007      & &  9.905(0.020) &  & 9.903(0.017)  & 18.749(0.014)  &  &                     &                      &    \\ 
               &          &         &             &&   9.95(0.05) &  &  9.95(0.05)    &  18.79(0.04)     &  &  9.94(0.04)   &  18.78(0.03)  &  $1^{-}$ \\
               &         &          &           && 9.995(0.007) &  & 9.995(0.007)    & 18.828(0.006)     & &  10.06(0.06)   &   18.88(0.05) &  $4^{+}$   \\
  10.10 &  7  &  0.10 & $-$0.003 && 10.097(0.040) & & 10.099(0.024)  & 18.917(0.021) & & 10.11(0.05) & 18.93(0.05) &   $2^{+},3^{-}$  \\ 
  10.29     &  72   &  0.08  &    0.034       && 10.270(0.007) && 10.271(0.007)   & 19.064(0.006) & &                       &                       &   \\
  10.35     & 88  &  0.20    &  $-$0.043   && 10.463(0.046) && 10.384(0.025)  & 19.161(0.021) & & 10.31(0.10) & 19.09(0.09) & $3^{-}$  \\
  10.60     &   3    &  0.05  &    0.002      & &10.643(0.034) &&  10.619(0.023)  & 19.362(0.019) &  &                      &             &                 \\ 
  10.83     & 110  &  0.10   &  $-$0.056   && 10.715(0.007) & & 10.721(0.007)  & 19.449(0.006) & &                      &               &                 \\    
  10.97     & 100  &   0.08  &     0.050     && 10.935(0.010) &&  10.939(0.009)  & 19.635(0.008) &  &                    &              &                  \\
  11.10     &  92  &   0.18  &    $-$0.048  && 11.113(0.047) & & 11.104(0.025)  & 19.777(0.022) & &                     &             &                \\
  11.30     &  48  &   0.03  &     0.015    & &11.273(0.038) & & 11.290(0.024)  & 19.936(0.020) & &  11.24(0.07) & 19.89(0.06) & $2^{+}$  \\
  11.35    &   6   &  0.07   & $-$0.003   && 11.400(0.017) & & 11.388(0.015)  & 20.020(0.013) & &  11.35(0.06)  & 19.99(0.05) &  $1^{-}$  \\ 
  11.48    &  42   &   0.05  &     0.021      && 11.465(0.030) & & 11.473(9.021)  & 20.092(0.018) & &              &               &                \\
  11.52    &  65   &   0.08  &   $-$0.035   && 11.585(0.007) &  &  11.582(0.007)  & 20.186(0.006) & &              &               &              \\
  11.73    & 125  &   0.07  &      0.065     && 11.750(0.024) & &  11.742(0.019)  & 20.323(0.016) & &             &             &                  \\ 
  11.90    & 220  &   0.04  &   $-$0.079   && 11.933(0.031) &  & 11.916(0.022)  & 20.472(0.018) & &               &                &                \\ 
  12.05    &   48  &   0.10  &    0.026      && 12.076(0.028) & &  12.064(0.020)  & 20.599(0.018) & & 12.09(0.13) & 20.62(0.11) & $4^{+}$ \\ 
  12.20    & 90    &   0.08  &  $-$0.050   && 12.200(0.017) & &  12.200(0.015)  & 20.715(0.013) & & 12.12(0.07) &  20.65(0.06) & $5^{-}$ \\ 
  12.30    &  55   &   0.10  &     0.030     & &12.343(0.037) & &  12.317(0.023)  & 20.815(0.020) & &            &                 &               \\
  12.50    &  45   &   0.07  &   $-$0.026  && 12.456(0.066) &  & 12.492(0.027)  & 20.966(0.023) & &              &                &                \\ 
              &        &            &                 &&  12.733(0.026) & &   12.733(0.026)    &  21.171(0.022)   & &               &                 &               \\   
              &        &            &          & & 12.840(0.017) & &   12.840(0.017)    &  21.263(0.015)   & & 12.81(0.05)   &    21.24(0.04)  &  $4^{+}$  \\   
  12.94    &  88  &   0.29   &     0.050    &&  12.960(0.010) &  & 12.958(0.009) & 21.364(0.008) & &            &                 &                \\ 
              &       &            &              && 13.110(0.024) &  &   13.110(0.024)  &   21.494(0.021)   & & 13.12(0.03)   &  21.51(0.02)    &  $3^{-}$ \\ 
  13.20\footnote{\hspace{1mm}continued in Table\,\ref{Fitparameter-2}} & 25 & 0.08 & $-$0.015 & &13.170(0.017) && 13.177(0.015) &  21.552(0.013) & &   13.15(0.09)   &  21.53(0.07)      &  $5^{-}$    \\\hline\hline \vspace*{-7mm}
\end{tabular}
\end{center}
\label{Fitparameter-1}
\end{table}
\begin{table}
\caption{Continuation of Table\,\ref{Fitparameter-1}.}
\begin{center}
\begin{tabular}{ccccccccccccc}\hline\hline
  1  &  2  &  3  &  4  &     &  5  &     &  6  &   7  &       &  8   &   9   &   10   \\
\multicolumn{4}{c}{present experiment}&~~~& $^{12}$C$(d,\alpha_{2})^{10}$B &~~~& & && \multicolumn{3}{c}{ $^{12}$C$(d,\alpha_{2})^{10}$B}      \\
$E_{0}$   & $\sigma_{\rm tot}$ & $\Gamma$ & $p_{\rm zz}$ & & $E_{\rm p}$ & & $\overline{E_{\rm d}}$ & $E(^{14}{\rm N}$*) &~~~~&$E_{\rm d}$ & $E(^{14}{\rm N}$*)  & $J^{\pi}$ \\
  (MeV)    &   (b$\cdot$MeV)      &   (MeV)      &         & & (MeV)  && (MeV)  &   (MeV) &  & (MeV) &     \\\hline
  13.30    &  35  &  0.10   &   0.021      & &  13.320(0.035)   & &  13.308(0.023)  &  21.664(0.019)  & &  13.32(0.05)  &  21.68(0.04) & $4^{+}$   \\
  13.50    &  40  &  0.04   &  $-$0.018  & &  13.490(0.008)   & &  13.491(0.008)  &  21.820(0.007)  & &                       &                    &                  \\
  13.59    &  38  &  0.04   &     0.023    & &  13.570(0.035)   & &  13.582(0.023)  &  21.898(0.019)  & &                       &                    &                 \\
  13.70    &  62  &  0.08   &  $-$0.036  & &  13.700(0.05)   & &   13.700(0.026) &  21.999(0.022)  & &                       &                     &                  \\
  13.76    &  20  &  0.04   &  0.009      & &   13.760(0.017)   & &  13.760(0.015)  &  22.051(0.013)   & &                      &                    &                  \\
  13.87    &  15  &  0.05   & $-$0.009   & &  13.813(0.025)   & &  13.836(0.019)  &  22.116(0.016)   & &                      &                    &                   \\
  13.96    &  19  &  0.06   &   0.011     & &   13.930(0.017)   & &  13.937(0.015)  &  22.202(0.013)   & &  14.00(0.03)  & 22.26(0.02) &  $4^{+}$    \\ 
              &        &           &                & &   14.067(0.115)   & &  14.067(0.115)     &  22.31(0.10)       & & 14.05(0.07)  &  22.30(0.06) &  $5^{-}$     \\ 
  14.45    & 297 & 0.30   &  $-$0.195   & &   14.413(0.043)  & &   14.438(0.025) &  22.631(0.021)    & &                    &                  &                  \\
  14.95   &   80  & 0.10   &   0.051       & &  14.861(0.109)   & &  14.944(0.029)  &  23.064(0.025)    & &                     &               &                   \\
  15.10   &   48  & 0.08   & $-$0.032    & &  15.000(0.030)   & &  15.050(0.021)  &  23.155(0.018)    & &                     &                &                    \\ 
  15.25   &   73  & 0.10   &    0.048     & &   15.203(0.050)   & &  15.238(0.026) &   23.315(0.022)    & &                     &                &                    \\
  15.40   &   50  &  0.10  &  $-$0.034  & &   15.440(0.055)   & &  15.409(0.026)&  23.462(0.023)    & & 15.33(0.07) & 23.40(0.06) &  $5^{-}$   \\
  15.62   & 134  &   0.10 &    0.088     & &   15.610(0.030)   & &  15.615(0.021)  &  23.638(0.018)   & &                      &                 &                     \\
  15.75   & 145  &  0.15  & $-$0.101   & &   15.700(0.030)   & &  15.725(0.021)  &  23.733(0.018)    & &                     &                &                     \\
  16.20   & 280  &  0.38  &  0.184       & &   16.210(0.066)   & &  16.202(0.027)  &  24.141(0.023)    & &                     &                &                    \\
  16.35   & 25    & 0.03  &  $-$0.011   & &                       & &  16.35(0.03)  &  24.268(0.026)    & &                     &                &                    \\
  16.50   & 39    &  0.10  &  0.027       & &  16.440(0.030)    & &  16.470(0.021)  &  24.370(0.018)    & &                    &                 &                   \\
  16.60   & 148  &  0.15  &  $-$0.107  & &  [16.650(0.030)]  & & 16.625(0.021)   & 24.503(0.018)     & &                    &                &                      \\
  16.86   &  23 &  0.11  &  0.016        & &   16.830(0.030)   & &  16.845(0.021) &   24.691(0.018)    & &                      &                 &                      \\
  17.16   & 163 &  0.30  &  $-$0.122   & &                       & &  17.16 (0.03)     &  24.96(0.03)       &  &                     &                 &                      \\
  17.18   &  75  &  0.15  &   0.054     &  &                        & &  17.18(0.03)     &  24.98(0.03)       & &                     &                &                      \\
  17.55   &  87  &  0.05  &   0.064     & &                         & &  17.55(0.03)     &  25.30(0.03)       & &                     &                &                      \\
  17.68   &  35  &  0.06 &  $-$0.026  & &                         & &  17.68 (0.03)    &  25.41(0.03)        & &                     &                &                       \\
  17.78   &  48  &  0.05 & 0.035        & &                         & &  17.78(0.03)      & 25.49(0.03)        & &                    &                 &                      \\
  17.87   &  55  &  0.08 & $-$0.042    & &                        & &   17.87(0.03)     & 25.59(0.03)        & &                    &                 &                     \\
  18.08   &  18  &  0.05 &  0.014       & &                         & &   18.08(0.03)     & 25.75(0.03)        & &                   &                 &                      \\
  18.14   &  15  &  0.05 &  $-$0.011  & &                        & &   18.14(0.03)     & 25.80(0.03)        & &                  &                 &                     \\
  18.23   &  53  &  0.05 &    0.039     &&                         & &   18.23(0.03)     & 25.88(0.03)       & &                  &                 &                     \\
  18.45   & 122 &  0.13 &  $-$0.096  & &                        & &    18.45(0.03)    &  26.07(0.03)       &  &                   &                &                    \\\hline\hline
\end{tabular}
\end{center}
\label{Fitparameter-2}
\end{table}
to three of Tables\,\ref{Fitparameter-1} and \ref{Fitparameter-2}. Under the assumption that the energies $E$ are uniformly distributed within the energy steps $\Delta E$=0.1\,MeV, the standard deviation of the fitted central energies $E_{0}$ is 0.1\,MeV/$\sqrt{12}$=0.035\,MeV. No errors are given to the $\sigma_{\rm tot}$ and  $\Gamma$ due to some arbitrariness in the choice of the parameters. Their values are to be regarded as adequate to describe the measured distribution.

Column four contains the $p_{zz}$ as they result from the single Gaussian fit functions in Eq.\,(\ref{PzzSum2}). Positive $p_{zz}$ are obtained by removal of deuterons from the beam in the $m=0$ state ($\sigma_{\pm 1}(E)=0$, $\sigma_{0}(E)>0$) and negative $p_{zz}$ by removal of deuterons in the $m=\pm 1$ state ($\sigma_{\pm 1}(E)>0$, $\sigma_{0}(E)=0$). With one exception, the fits yield a sequence of alternating signs. The statistical accuracy and the 0.1\,MeV step width are insufficient to resolve further adjacent $\Delta p_{zz}(E)/\Delta E$ ranges of the same sign.

Astonishing agreement was observed between central energies $E_{0}$, resulting from the fit, and energies of resonances in the reaction $^{12}{\rm C}(d,\alpha_{2})^{10}$B with population of the second excited state in $^{10}$B given in an earlier data compilation of\,\cite{Ajz-Sel_1991} and the actual one\,\cite{NNDC} with reference to\,\cite{Jolivette_1981}. The reaction is understood as including intermediate $^{14}$N states. The deuteron resonance energies, the corresponding energies of exited $^{14}$N states, and spin/parity of these states are given in columns 8, 9, and 10 of the tables. They were obtained by an S-matrix analysis of measured angular distributions\,\cite{Smith_1972,Jolivette_1974}. Unfortunately, the documents with the comprehensive data on the measured  excitation functions, given as reference in Ref.\,\cite{Smith_1972}, are no longer available\,\cite{ZB}. Therefore, the peak positions had to be read from the excitation functions of Figs. 2 to 5 of\,\cite{Smith_1972} as it had to be done for additional data from\,\cite{Ehrenstein_1971} (there Fig.\,2) and \cite{Jolivette_1974} (there Fig.\,4). The read peak energies are given in column 5 of the tables. Their errors base on the bin widths 0.05\,MeV\,\cite{Smith_1972},  0.10\,MeV\,\cite{,Jolivette_1974}, and 0.10\,MeV\,\cite{Ehrenstein_1971}, the quality of the peaks, and their frequency in the figures. Column 6 contains the weighted averages with errors resulting from the present central energies $E_{0}$ (column1) and the peak energies $E_{\rm p}$ (column 5). In column 7 one finds the corresponding $^{14}$N excitation energies.

From the 40 fitted central energies $E_{0}$ in the range 9.55 to 16.86\,MeV, with the exception of that at 16.35\,MeV 39 have a counterpart in the $(d,\alpha_{2})$ peak positions. The distribution of the 39 differences  $E_{\rm p}-E_{0}$ follow a Gaussian distribution with the center at $-$0.009\,MeV and a standard deviation $\sigma$=0.035\,MeV. The reduced $\chi^{2}=1/38*\sum_{i=1}^{39}(E_{0,i}-E_{\rm p,i})^{2}/(\Delta E_{0,i}^{2}+\Delta E_{\rm p,i}^{2})=0.98$ confirms the error estimates. Column six presents the weighted averages $\overline{E}_{\rm d}$ of $E_{0}$ and $E_{\rm p}$ and their errors. The narrow sequence of peaks at $E_{\rm p}$=9.625, 9.640, and 9.665\,MeV and the peaks at 9.95 and 9.995\,MeV in the $(d,\alpha_{2})$ reaction (column five) cannot be resolved in the present $p_{zz}$ fluctuations. The peaks at $E_{\rm p}$=12.733, 12.840, and  13.110\,MeV lie in the gap between the C58 and C94 data, where the energy dependence of $p_{zz}$ is covered by a single fitted Gaussian. The peak at $E_{\rm p}$=14.067\,MeV is an additional peak in the gap between the C94 and C129 data which is covered by the broad  Gaussian with $E_{0}$=14.45\,MeV.

 Part of the levels above 24.7\,MeV, resulting from the present analysis only, and also levels below are found as peaks in the $^{10}{\rm B}(\alpha,d)^{12}{\rm C}$ excitation spectra\,\cite{Al'-dzhauakhiri_1968} and as components in the wide distributions of the $^{10}{\rm B}(\alpha,p)^{13}{\rm C}$ excitation spectra\,\cite{Spasskii_1966}. Especially the highest level at 26.07$\pm$0.03\,MeV with a large removal cross section is confirmed by distinct peaks in the $^{10}{\rm B}(\alpha,d)^{12}{\rm C}$ excitation functions at $E_{\alpha}= 20.21\pm 0.02$\,MeV or $E(^{14}{\rm N}^{*})=26.04\pm 0.02$\,MeV.
\section{Discussion of the fit results~\label{Discussion}}
The fit results are discussed under three aspects,
\begin{itemize}
\item the coarse structure and the $^{14}$N giant dipole resonance (Sec.\,\ref{Coarse structure}),
\item the rapid fluctuations and the energy dependence of $d-^{12}$C reactions (Sec.\,\ref{Fine structure}), and
\item the signs of $\Delta p_{zz}/\Delta E$ and the parity of $^{14}$N states (Sec.\,\ref{Signs}).
\end{itemize}
\subsection{Coarse structure~\label{Coarse structure}}
The production of tensor polarization $p_{\rm zz}$ due to the quadrupole deformation of the deuteron with smooth dependence on the areal carbon-target density\,\cite{Baryshevsky+Rouba_2010}, discussed in the introduction (Sec.\,\ref{Introduction}), is superposed by essentially stronger nuclear structure effects.

The central energy of the (negative) peak in the energy dependence of $p_{zz}$ (Fig.\,\ref{Pzz2}) at 15.3\,MeV corresponds to a $^{14}$N excitation energy of 23.4\,MeV, which is in agreement with the energy of the $^{14}$N giant dipole resonance (GDR), spreading around 23.4\,MeV\,\cite{Berman+Fultz_1975}. The other characteristic parameter of the GDR is its width. In the collective model for spherical nuclei the GDP was described as oscillation of the bulk of protons against the bulk of neutrons\,\cite{Goldhaber+Teller_1948} or as hydrodynamic flow of the neutrons against the protons under conservation of the nucleon density\,\cite{Steinwedel+Jensen_1950}. In the extension to deformed nuclei of ellipsoidal shape, two oscillation modes appear $-$ a slower with motion along the major axis (half axis $a$, projection quantum number $m=\pm 1$, oscillation energy $E_{\rm a}=\hbar \omega_{a}$) and a faster along the minor axis (half axis $b$, $m=0$, $\hbar \omega_{b}$)\,\cite{Danos_1958,Okamoto_1958}. For prolate cigar-shaped deformation with $a>b$ and $E_{b}>E_{a}$ the eigenvalues of the hydrodynamic model yield the relation\,\cite{Danos_1958}
\begin{equation}
\frac{E_{b}}{E_{a}}=0.911\cdot \frac{a}{b}+0.089.
\label{Def}
\end{equation}
The values of the half axes $a$ and $b$, needed to calculate the ratio of the oscillation energies, are obtained from two relations.. The intrinsic quadrupole moment is $Q_{0}=2/5\cdot Z\cdot(a^{2}-b^{2})$ with $Z=7$ (e.g.\,\cite{Mayer-Kuckuk}).Furthermore, under assumption of constant nuclear mass density  the volume $4\pi ab^{2}/3$ of the prolate, spheroidal nucleus is equal to the volume $4\pi R_{0}^{3}/3$ of the spherical nucleus of radius $R_{0}$, i.e., $R_{0}^{3}=ab^{2}$ or $b^{2}=R_{0}^{3}/a$. The two relations yield the cubic equation for the half axis $a$,
\begin{equation}
a^{3}-\frac{5Q_{0}}{2}\cdot a-R_{0}^{3}=0.
\label{Cubic_a}
\end{equation}
In the ground state $^{14}$N is of prolate deformation with $Q_{0}=(1.93\pm 0.08)$\,fm$^{2}$\,\cite{Ajz-Sel_1991}. As no data were found for the excited states around 23\,MeV, the assumption is made that the excited nucleus oscillates around the ground-state shape. Two values for $R_{0}$ are available, (1) the standard nuclear radius $R_{0}$=1.2\,fm$\cdot 14^{1/3}$=2.89\,fm and (2) the rms charge radius $<r_{\rm c}^{2}>^{1/2}$=2.56\,fm\,\cite{Schaller_1980} given in\,\cite{Ajz-Sel_1991}. The solution of Eq.\,(\ref{Cubic_a}) for (1) yields $a$=2.97\,fm and $b$=2.85\,fm. The ratio $a/b$=1.043 yields $E_{b}/E_{a}$=1.039. For (2) one obtains $a$=2.65\,fm, $b$=2.52\,fm,  $a/b$=1.055, and $E_{b}/E_{a}$=1.050. 

The two ratios $E_{b}/E_{a}$ are compared with the ratio of the average $^{14}$N excitation energies $\overline{E}_{\rm b}(^{14}$N*) above and $\overline{E}_{\rm a}(^{14}$N*) below the $p_{zz}$ peak center at 23.4\,MeV corresponding (Fig.\,\ref{Pzz2}) to the deuteron-peak energy $E_{\rm d}$=15.3\,MeV. The average values are calculated as $\overline{E}_{\rm a,b}(^{14}{\rm N}^{*})=\sum E(^{14}{\rm N}^{*})\cdot S(E(^{14}{\rm N}^{*}))/\sum S(E(^{14}{\rm N}^{*}))$. The $^{14}$N excitation spectrum $S(E(^{14}{\rm N}^{*}))$ is calculated with the use of the 51 fitted Gaussian-shaped removal cross sections. For $\overline{E}_{\rm a}$ the summation runs from 22.3$\pm$0.2 ($E_{\rm d}=14.1\pm 0.2$\,MeV) to 23.4\,MeV, for  $\overline{E}_{\rm b}$ from 23.4 to 24.4$\pm$0.2\,MeV ($E_{\rm d}=16.5\pm 0.2$\,MeV).The resulting $\overline{E}_{\rm b}(^{14}$N*)=23.91$\pm$0.09\,MeV  and $\overline{E}_{\rm a}(^{14}$N*)=22.86$\pm$0.03\,MeV yield the ratio $\overline{E}_{\rm b}(^{14}$N*)/$\overline{E}_{\rm b}(^{14}$N*)=1.046$\pm$0.004. Within the error it is compatible with both 1.039 and 1.055 given above and obtained with use of the relation of Eq.\,(\ref{Cubic_a}) from the hydrodynamic model and the assumption of equal nucleon density in a spherical and in a deformed, oscillating nucleus.

For $E_{\rm in}$ in the range from about 14\,MeV to the peak center at 15.3\,MeV $\Delta p_{zz}/\Delta E$ on the average is negative. In the fit, $\Delta p_{zz}/\Delta E <0$ is caused by coupling deuterons in the $m=\pm 1$ state to the $^{12}$C nuclei. This agrees with the theoretical prediction that at the lower energy $E_{a}$ the oscillations are along the major axis with projection quantum number $m=\pm 1$. On the other hand, above the peak center up to about 16.5\,MeV $\Delta p_{zz}/\Delta E$ on the average is positive. This is described by coupling of deuterons in the $m=0$ state in agreement with the theoretical prediction that at the higher energy $E_{b}$ the oscillations are along the minor axis with projection quantum number $m=0$.

In an earlier experiment, the energy-dependent removal of deuterons from a beam was utilized to study the energy dependence of the ${^{12}\rm{C}}(d,n){^{13}\rm{N_{g.s.}}}$ cross section\,\cite{Wilkinson_1955}. A 20.5\,MeV deuteron beam was slowed down to about 5\,MeV  in a stack of 28 polyethylene foils of 0.1\,mm thickness. The $\beta ^{+}$ activity of ${^{13}\rm{N}}$ (half life 9.96\,min) produced in each of the foils was measured. The cross section was reported\,\cite{Wilkinson_1955} as increasing smoothly with decreasing deuteron energy without resonances, which had been observed at low deuteron energies. The energy dependence of the cross section, however, shows a wide-spread enhancement for those foils, in which the deuteron energies were between about 14 and 16.5\,MeV. The range coincides with that of the (negative) peak in $p_{zz}$, which is related to the excitation of the $^{14}$N giant dipole resonance. This indicates that in the ${^{12}\rm{C}}(d,n){^{13}\rm{N_{g.s.}}}$ reaction,too, the formation of intermediate excited $^{14}$N states leads to the enlargement of the reaction cross section. Besides direct proton pick-up, this possibility was discussed in\,\cite{Wilkinson_1955} without, however, a definite conclusion.

The full width of the (negative) peak in Fig.\,\ref{Pzz2} is about 2.4\,MeV. According to the uncertainty relation $\Delta E\cdot \Delta t=\hbar$ the corresponding time spread is $3.3\cdot 10^{-22}$\,s. This about twice the time of $1.7\cdot 10^{-22}$\,s a deuteron of 15\,MeV ($v_{\rm cms}=3.2\cdot 10^{9}$\,cm/s) needs to pass a $^{12}$C nucleus of $2\cdot 1.2\cdot 10^{-13}\cdot 12^{1/3}=5.5\cdot 10^{-13}$\,cm diameter.
\subsection{Fine structure~\label{Fine structure}}
Deuteron removal from the beam is caused by all elastic and inelastic $d-^{12}$C interactions. In general, the reactions  are characterized by a weak energy dependence or wide resonances in the excitation functions including the isospin-allowed (0+0$\rightarrow$0+0) reaction channels $^{12}{\rm C}(d,\alpha_{0,1,3})^{10}{\rm B}$ with population of the isopin-zero ground, first and third excited states in $^{10}$B. One would expect that their contributions to the removal of deuterons from the beam cover the much weaker contribution by the isospin-hindered  (0+0$\rightarrow$0+1) $^{12}{\rm C}(d,\alpha_{2})^{10}{\rm B}$ reaction with population of the second excited $J^{\pi}=0^{+},T=1$ state at 1.74\,MeV in $^{10}$B. The fit to the fluctuations in $p_{zz}$, however, yields central energies $E_{0}$ of the removal cross sections, which are in astonishing agreement with the energies $E_{\rm p}$ of the narrow resonances in the $(d,\alpha_{2})$ channel. As mentioned above, from the 40 fitted central energies $E_{0}$ in the range 9.55 to 16.86\,MeV, 39 have a counterpart in the $(d,\alpha_{2})$ peak positions. The further nine $(d,\alpha_{2})$ peaks without $E_{0}$ counterparts lie in the energy range difficult to fit ($E_{\rm p}$=9.625 to 9.995\,MeV) and in the energy gaps of the present measurement ($E_{\rm in}$=12.50 to 13.00 and 14.00 to 14.80\,MeV).

Due to the isospin characteristics $0+0\rightarrow 0+1$, the reaction $^{12}{\rm C}(d,\alpha_{2})^{10}$B with population of the second excited $J^{\pi0}=0^{+},T=1$ state at 1.74\,MeV in $^{10}$B would be isospin-forbidden. The isospin impurity of the deuteron ground state is expected to be negligible and those of the $^{12}$C ground state, the $\alpha$ particle, and the second excited $^{10}$B state were estimated as $1\times 10^{-3}$, $1\times 10^{-5}$, and $7\times 10^{-4}$, respectively\,\cite{MacDonald_1960}. Isospin-1 admixture in the $\alpha$ particle caused by the $\alpha-^{10}$B Coulomb interaction was estimated as $2\times 10^{-3}$\,\cite{Griffy_1966}. The measured ratio of the isospin-forbidden $\alpha_{2}$ total cross section to those of the isospin-allowed $\alpha_{0}$, $\alpha_{1}$, and $\alpha_{3}$ channels of $\approx$1\% is by an order of magnitude larger than these impurities and also by an order of magnitude larger than predicted for single- or multi-step direct-reaction mechanisms\,\cite{Smith_1972}. Intermediate isospin-mixed $^{14}$N states were regarded as a way to explain the breaking of the isospin conservation. 

\begin{figure}
\begin{center}
\includegraphics[width=8cm]{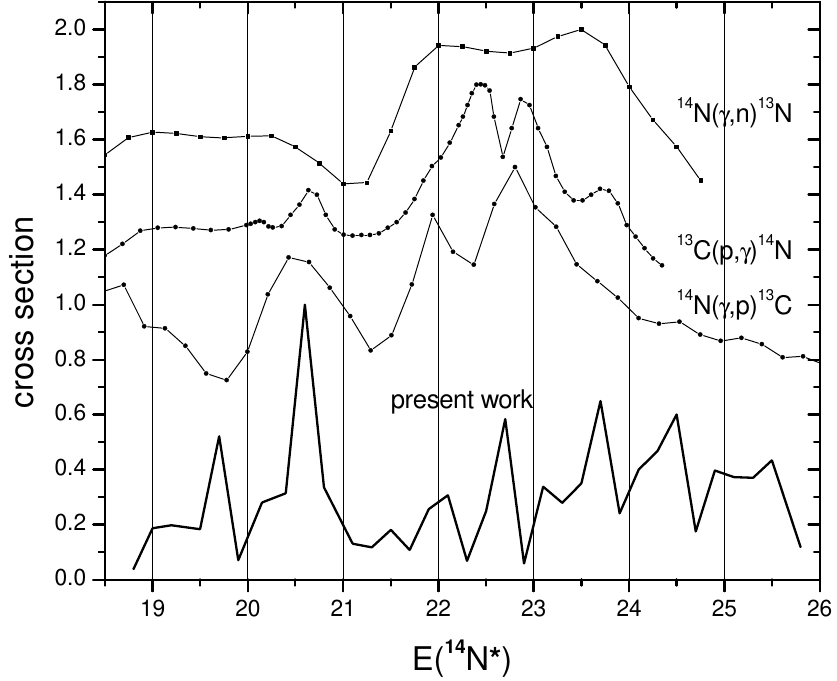}
\caption{The $^{14}$N excitation spectrum resulting from the energy dependence of the cross sections describing removal of deuterons from the beam (bin width 0.2\,MeV) compared with excitation functions measured for the reactions $^{14}{\rm N}(\gamma,p)^{13}{\rm C}$\,\cite{Kosiek_1964}, $^{13}{\rm C}(p,\gamma)^{14}{\rm N}$\,\cite{Riess_1971}, and  $^{14}{\rm N}(\gamma,n)^{13}{\rm C}$\,\cite{King_1960}. All four functions are shown with the highest peak normalized to one. The functions for $(\gamma,p)$, $(p,\gamma)$, and $(\gamma,n)$ for readability are shifted up by 0.5, 0.8, and 1.0, respectively.}
\label{EXC}
\end{center}l
\end{figure}

The agreement of the $(d,\alpha_{2}) $ peak energies and the fitted central energies $E_{0}$ indicates that deuteron removal from the beam leads to formation of excited $^{14}$N states. In case that this happens, the structure of the resulting $^{14}$N excitation spectrum should be similar to the excitation functions of reactions with excitation and de-excitation of $^{14}$N states. The sum of the 51 Gaussian cross section fit functions yields the $^{14}$N excitation spectrum with the energy relation $E(^{14}$N)=0.856$\cdot E$+10.272 (energies in MeV). The first term yields the center-of-mass deuteron energy and the second accounts for the $d+^{12}$C$-^{14}$N mass balance\,\cite{Ajz-Sel_1991}. As Fig.\,\ref{EXC} shows, the peaks in the present excitation spectrum find counterparts in the other excitation functions. This conformity 
and the agreement of the fitted central energies with the peak energies in the $(d,\alpha_{2})$ reaction support the finding that identical $^{14}$N states are created in both cases.

Spin and parity $J^{\pi}$ of the intermediate $^{14}$N states in the reaction $d+^{12}{\rm C}\rightarrow ^{14}{\rm N}^{*}\rightarrow \alpha +^{10}$B$^{*}$(1.74\,MeV,$J^{\pi}=0^{+},T=1$) are determined by the spin/parity relation $1^{+}+0^{+}\rightarrow J^{\pi} \rightarrow 0^{+}+0^{+}$ and the initial and final orbital angular momenta $l_{\rm i}$ and $l_{\rm f}$. Parity conservation requires $l_{\rm i}=l_{\rm f}=l$ and $l=0$ is excluded. Furthermore for the intermediate $^{14}{\rm N}^{*}$ $J=l_{\rm f}=l$ and $\pi=(-1)^{l}$. The initial orbital momentum vector is perpendicular to the direction of the incident beam as z axis. Therefore $l_{i,z}=0$ and $J_{z}=m$, where $m=0, \pm 1$ is the z component of the deuteron spin. For $l_{z}=0$, $m=0$, $J_{z}=m=0$, and $J=l$ the Clebsch-Gordan coefficients $(l\hspace{0.5mm}s\hspace{0.5mm}  l_{z }\hspace{0.5mm} s_{z}\hspace{-1mm}\mid\hspace{-1mm} J \hspace{0.5mm}J_{z})=(l\hspace{0.5mm}s\hspace{0.5mm}  l_{z }\hspace{0.5mm} s_{z}\hspace{-1mm}\mid\hspace{-1mm} l \hspace{0.5mm}s_{z})=(l\hspace{0.5mm}1\hspace{0.5mm} 0\hspace{0.5mm} 0\hspace{-1mm}\mid\hspace{-1mm} l \hspace{0.5mm}0)=0$. Therefore, contributions by deuterons in the $m=0$ state were excluded in the S-matrix fit by selection of $l_{z}=\pm1$ for outgoing waves\,\cite{Jolivette+Richards_1969,Jolivette_1981}. Variation of the S-matrix elements (resonance energies, widths, complex product of the initial and final partial widths, parameters of a complex background function) in the $\chi^{2}$ fit to the angular distributions of\,\cite{Smith_1972,Jolivette_1974} lead to the states given in the right-hand columns of Tables\,\ref{Fitparameter-1} and \ref{Fitparameter-2}.

When one regards the formation only of the $^{14}$N states, due to the perpendicular orientation of the initial orbital angular momentum vector against the beam direction as z axis the condition $l_{z}=0$ is maintained. The limitation by $l_{z}=\pm 1$, imposed by the exit channel in the $(d,\alpha_{2})$ reaction, drops. Contrary to the $(d,\alpha_{2})$ case, where intermediate $^{14}$N states with $J=l$ only are allowed, the restriction to formation allows $^{14}$N states with $J=l$ as well as $J=\pm 1$, all of parity $(-1)^{l}$. The fit by Gaussian functions to the ranges of positive or negative $\Delta p_{zz}(E)/\Delta E$ in the observed fluctuations of $p_{zz}$, however, yields central deuteron energies, which are in excellent agreement with the energies of peaks in the $(d,\alpha_{2})$ excitation functions. This leads to the conclusion that in the resonance-shaped removal of 9.5 to 18.6\,MeV deuterons from the beam the formation of those $^{14}$N states dominates, which with $J=l$ are the allowed intermediate states in the $d+^{12}{\rm C}\rightarrow ^{14}{\rm N}^{*}\rightarrow \alpha +^{10}$B$^{*}$(1.74\,MeV,$J^{\pi}=0^{+},T=1)$ reaction.
\subsection{Signs of $p_{zz}$ and parities of $^{14}$N states~\label{Signs}}
The errors of $\overline{E}(^{14}$N*) in column seven of Tables\,\ref{Fitparameter-1} and \ref{Fitparameter-2} and those of the level energies from the $\chi^{2}$ fit in column nine, estimated\,\cite{Jolivette_1981,Ajz-Sel_1991} as about 10\% or 20\% of the calculated level widths, allow ambiguities in associating the levels of column nine with those of column seven. The arrangement in the tables is the most reasonable one regarding the energies and errors of adjacent levels.

From the 20 levels, obtained in the S-matrix fit, 12 within the errors have a counterpart among the levels of column seven with $p_{zz}$ obtained from the present measurement (column 4). Among the 8 levels without counterpart, 5 lie in the deuteron-energy range up to 10\,MeV, where the weak fluctuations restrain a reasonable fit. The calculated 2 levels at 21.24 and 21.51\,MeV lie in the gap between the present C58 and C94 data (Fig.\,\ref{Pzz1}), that at 22.30\,MeV in the gap between the C94 and the C129 data. For the 11 levels with partner, the sign of $p_{zz}$ is found in surprising agreement with that of the assigned parity. For 6 levels the negative parity agrees with the negative sign of $p_{zz}$ and for 5 levels the positive parity is in agreement with the positive sign of $p_{zz}$. For the 18.93$\pm$0.05\,MeV level differing spin/parity assignments are given, $2^{+}$ in the original work\,\cite{Jolivette_1981} and $2^{+},3^{-}$ in the data compilation\,\cite{Ajz-Sel_1991}. The negative $p_{zz}=-0.003$ of the 18.917$\pm$0.021\,MeV level would contradict the assignment $2^{+}$ and would leave $3^{-}$.

The calculated 22.26$\pm$0.02\,MeV $J^{\pi}=4^{+}$ level is a special case. It resulted from the $\chi^{2}$ fit\,\cite{Jolivette_1981} with an exceptionally small width and was commented as "not obvious in the data but substantially improving the fit". The deuteron energies $E_{0}$=13.96$\pm$0.035\,MeV and $E_{\rm p}$=13.930$\pm$0.017\,MeV yield a $^{14}$N level at $E(^{14}$N*)=22.202$\pm$0.013\,MeV. The positive $p_{zz}$=+0.011 would require positive parity. Thus, the postulated 22.26$\pm$0.02\,MeV, $J^{\pi}=4^{+}$ level is associated with the 22.202$\pm$0.013\,MeV level. According to the errors, the 22.30$\pm$0.06\,MeV $J^{\pi}=5^{-}$ level would be associated with the 22.31$\pm$0.10\,MeV level from the $(d,\alpha_{2})$ data. In the present measurements, it lies in the gap between the C94 and the C129 data. The fit, covering the wide gap between the deuteron energies 14.00 and 14.80\,MeV with negative $\Delta p_{zz}(E)/\Delta E$, is obtained with $E_{0}$=14.45\,MeV. In Table \,\ref{Fitparameter-2} due to the agreement in energy the level is connected with the 14.413$\pm$0.043\,MeV level from the $(d,\alpha_{2})$ reaction. Within the gap of the present data, however, it would be possible to associate the negative $p_{zz}=-0.195$ with the $(d,\alpha_{2})$ peak at 14.07$\pm$0.12\,MeV. The negative slope in the gap would be centered at this energy corresponding to a 22.31$\pm$0.10\,MeV $^{14}$N level with a negative $p_{zz}$ in agreement with the negative parity of the calculated 5$^{-}$ level at 22.30$\pm$0.06\,MeV level. The level at 22.631\,MeV would contribute to the negative slope or it could be associated with a positive $p_{zz}$.\vspace{3mm}

Excited $^{14}$N states above 18.4\,MeV are given in\,\cite{Ajz-Sel_1991} and\,\cite{NNDC} in addition to those from the $(d.\alpha_{2})$ reaction only. These and further levels from other reactions, not given there, can be used to check the relation between their parity and the signs of $p_{zz}$.

\underline{20.11$\pm$0.02\,MeV} with $J^{\pi}=3^{-}, 4^{-}$: it can be associated with the 20.159$\pm$0.018\,MeV level from $(d,\alpha_{2})$ and the present work. The negative $p_{zz}=-0.035$ would agree with negative parity and exclude $4^{-}$ due to the requirement of natural parity.

\underline{21.8\,MeV} with $J^{\pi}=4^{+}$: the width $\Gamma \sim$0.8\,MeV covers about eight levels with positive and negative $p_{zz}$, which does not allow a conclusion about the sign relation.

\underline{22.5\,MeV and 23.0\,MeV} both with $J^{\pi}=2^{-}$: the strong resonances in the reaction $^{13}{\rm C}(p,\gamma_{0})^{14}{\rm N}$ with the $\gamma$ transition to the $^{14}$N $J^{\pi}=1^{+}$ ground state are understood as $J^{\pi}=2^{-}$ giant dipole resonance states in $^{14}$N\,\cite{Riess_1971}. In the excitation-energy range around 23\,MeV, however, also $0^{-},1^{-}$ levels may exist\,\cite{Paul_1975}. The negative parity would agree with the large negative $p_{zz}=-0.195$, which results from the fit to $\Delta p_{zz}(E)/\Delta E<0$ in the wide gap between the C94 and the C129 data. The lack of measured $p_{zz}$ data between those for deuteron energies of 14.00 and 14.80\,MeV, corresponding to $^{14}$N excitation energies between 22.26 and 22.94\,MeV, and the $(d,\alpha_{2})$ data do not allow the conclusion whether states of unnatural parity like those with  $J^{\pi}=2^{-}$, too,  are created in the removal of deuterons from the beam. In any case, the negative $p_{zz}$ agrees with the parity of the two states.

\underline{23.7\,MeV} not given in\,\cite{Ajz-Sel_1991}: A resonance centered at 23.7\,MeV with a width $\Gamma \sim$0.6\,MeV was found in the $(p,\gamma_{0})$ and $(p,\gamma_{1})$ excitation functions with $\gamma$ emission to the $1^{+},T=0$ $^{14}$N ground state and the first excited $0^{+},T=1$ state at 2.31\,MeV\,\cite{Riess_1971}. A "very tentative" $J^{\pi}=1^{-}$ assignment was made to the 23.7\,MeV level. Negative parity is supported by the observation of a peak in the $E1$ photo-proton production\,\cite{Baglin_1974} on the  $^{14}$N $1^{+}$ ground state. The 23.7\,MeV resonance can be associated with the 23.754\,MeV level of Table\,\ref{Fitparameter-2}. The large, negative $p_{zz}=-0.101$ agrees with the negative parity assignment and $J^{\pi}=1^{-}$ would fulfill the requirement of natural parity.  The narrow adjacent resonance at 24.15\,MeV in the $(p,\gamma_{1})$ excitation function\,\cite{Riess_1971} was not confirmed in a later work\,\cite{Paul_1975}.

\underline{24.0\,MeV}: A broad resonance, spreading from about 24 MeV to about 25.5 MeV $^{14}$N excitation energy, is found in the $(p,\gamma_{0})$ and $(p,\gamma_{1})$ excitation functions\,\cite{Paul_1975}. Its presence in the $^{14}{\rm N}(\gamma,n_{0})^{13}{\rm N}$\,\cite{Jury_1980} excitation functions allows $J^{\pi}=0^{-}, 1^{-}$ and $2^{-}$. The two energies $E_{0}$=16.60 and 17.16\,MeV correspond to levels at 24.48 and 24.96\,MeV, which both lie in the energy range of this resonance. The  large negative $p_{zz}=-0.107$ and $-0.122$ agree with the negative parity assigned to the resonance. A fragmentary $(d,\alpha_{2})$ excitation function with deuteron energies up to 17.0\,MeV\,\cite{Ehrenstein_1971}, lets space for a peak at a deuteron energy $E_{\rm p}$=16.65$\pm$0.10\,MeV. Together with $E_{0}=16.60\pm 0.03$\,MeV it would yield an excited level at 24.485$\pm$0.025\,MeV as intermediate state in this reaction. The second $E_{0}$=17.16\,MeV is above 17.0\,MeV. A further $(d,\alpha_{2})$ excitation-function measurement\,\cite{Jaenecke_1968} gives cross sections at deuteron energies 16.0, 17.0, and 18.0\,MeV. The value at 17.0\,MeV is slightly larger than those at 16.0 and 18.0\,MeV, which indicates the population of an intermediate $^{14}$N state around 24.8\,MeV. If both levels at 24.485 and 24.8\,MeV are intermediate $^{14}$N states in the $(d,\alpha_{2})$ reaction, this excludes $J^{\pi}=0^{-}$ and $2^{-}$ and leaves $1^{-}$ with natural parity.

\vspace{3mm}The 0.1\,MeV energy steps and the statistical errors of the present measurement do not allow to resolve narrow and weak peaks in the ranges of positive or negative $\Delta p_{zz}(E)/\Delta E$ in addition to those given in  Tables\,\ref{Fitparameter-1} and \ref{Fitparameter-2}.. The overall sign in the intervals, however, would be maintained. The sequence of $^{14}$N states of alternating positive and negative parity would confirm the interpretation of the asymmetry observed in the angular distributions in the  $^{12}{\rm C}(d,\alpha_{2})^{10}{\rm B}$ reaction\,\cite{Smith+Richards_1969,Ehrenstein_1971,Smith_1972}.
\section{Possibility to produce tensor-polarized deuteron beams~\label{Application}}
\begin{figure}[h]
\begin{center}
\includegraphics[width=8cm]{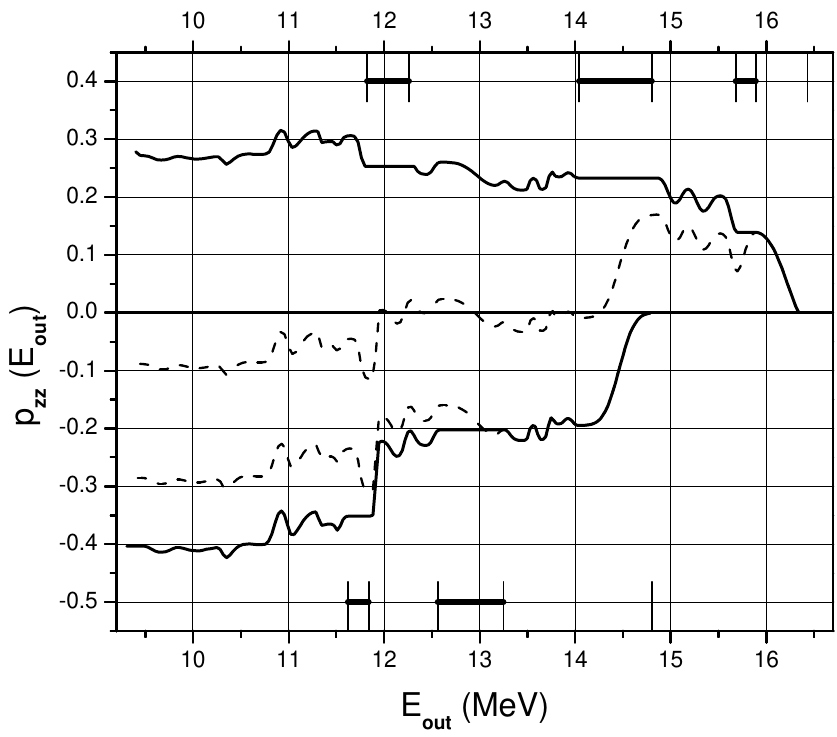}
\caption{Development of the tensor polarization $p_{\rm zz}$ calculated with Eq.\,(\ref{Pzz(E)}) in slowing-down from two initial energy $E_{\rm in}$ to $E_{\rm out}$ in the target. {\bf Upper part}: the dashed line gives $p_{\rm zz}(E_{\rm out})$ during slowing down from $E_{\rm in}$=16.4\,MeV in a pure carbon target. The full line gives the development in a sandwich target with layers of non-polarizing material inserted to slow down in the energy ranges, given by the upper bars, to bridge the ranges of negative  $p_{\rm zz}$ production. {\bf Lower part}: the corresponding curves for $E_{\rm in}$=14.8\,MeV.}
\label{PzzProduction}
\end{center}
\end{figure}

The present results indicate the possibility to produce a tensor-polarized, forward-transmitted deuteron beam from an initially unpolarized beam by the $d$-$^{12}$C interaction in carbon targets. The development of the tensor-polarization component $p_{zz}$ during slowing-down of an initially unpolarized beam in a carbon target from the initial energy $E_{\rm in}$ to an energy $E_{\rm out}$ behind the target can be calculated with the use of a modified version of Eq.\,(\ref{PzzSum2}),
\begin{eqnarray}
p_{\rm zz}(E_{\rm out})=\frac
  {2\cdot{\rm exp}\,\{-\rho \sum_{n=1}^{{\rm n}(E_{\rm out})} \sum_{{\rm i}=1}^{25} \sigma_{\pm 1,{\rm i}}(E_{\rm n})\Delta x\}
  -2\cdot{\rm exp}\,\{-\rho \sum_{n=1}^{{\rm n}(E_{\rm out})} \sum_{{\rm j}=1}^{24}       \sigma_{0,{\rm j}}(E_{\rm n})\Delta x\}}
  {2\cdot{\rm exp}\,\{-\rho \sum_{n=1}^{{\rm n}(E_{\rm out})} \sum_{{\rm i}=1}^{25} \sigma_{\pm 1,{\rm i}}(E_{\rm n})\Delta x\}
   +{\rm exp}\,\{-\rho \sum_{n=1}^{{\rm n}(E_{\rm out})} \sum_{{\rm j}=1}^{24}        \sigma_{0,{\rm j}}(E_{\rm n})\Delta x\}},
\label{Pzz(E)}
\end{eqnarray}
where the outer sums now run from n=1 ($E_{\rm n}\approx E_{\rm in}$) to the layer where $E_{\rm n}\approx E_{\rm out}$. The cross sections are given by the parameters of Tables\,\ref{Fitparameter-1} and\,\ref{Fitparameter-2}. Those with $p_{zz}>0$ in column 4 are attributed to $\sigma_{0,{\rm i}}(E_{\rm n})$ and those with $p_{zz} < 0$  to $\sigma_{\pm 1,{\rm i}}(E_{\rm n})$. Figure\,\ref{PzzProduction} shows the calculated development of $p_{zz}(E_{\rm out})$ for the two initial energies 16.3 and 14.8\,MeV. The upper dashed line shows that slowing down from 16.4\,MeV to $E_{\rm d}$=14.8\,MeV in a pure carbon target of areal density 30.6\,mg/cm$^{2}$ yields a 
beam behind the target with $p_{\rm zz}=+0.17$. For lower $E_{\rm d}$ $p_{\rm zz}$ approaches zero due to the production of negative $p_{\rm zz}$ below 14.8\,MeV. This can be suppressed by bridging the energy range from 14.8 to 14.04\,MeV by slowing down in a non-polarizing material of appropriate areal density. As it is shown by the upper full curve of Fig.\,\ref{PzzProduction} with two additional bridging ranges, $p_{\rm zz}$ values around +0.3 can be achieved for the whole $E_{\rm d}$ range below 15\,MeV. For $E_{\rm in}$=14.8\,MeV even with a pure carbon target $p_{\rm zz}$ values reaching $-$0.3 are achieved which can be enlarged to $-$0.4 by two bridging ranges.

In earlier measurements, the tensor polarization of deuteron beams was measured with the use of the polarimeter reaction  $^{3}{\rm He}(\vec{d},p)^{4}{\rm He}$ in setups similar to the present one. To get the reaction induced by s-wave deuterons, the deuteron beams were slowed down from energies up to 13\,MeV down to 1.5\,MeV in beryllium foils\,\cite{Goddard_1976}, from up to 7.5\,MeV to 0.8\,MeV in polyethylene foils\,\cite{Seiler_1964}, and from up to 6.8\,MeV to below 1\,MeV in mylar foils\,\cite{Cords_1969_2}. This would allow one to produce tensor-polarized deuteron beams with $p_{\rm zz}$ around +0.3 or $-0.4$ of energies down to about 0.8\,MeV by adding graphite or foils of the mentioned materials behind the sandwiched carbon target. 

The process of energy loss in a target is, however, connected with angle and energy straggling. The widths of the angular and the energy distribution both increase with increasing target thickness. Therefore thin targets might be preferable. The use of a 9.60\,mg/cm$2$ carbon foil to slow down a 16.35\,MeV beam to 15.89\,MeV according to Fig.\,\ref{PzzProduction} would yield a beam with $p_{\rm zz}=+0.14$. The widths  of the angular and the energy distributions (full-widths-half-maximum of Gaussian distributions) would be $\Gamma_{\rm angle}=0.81\,^{\circ}$ and $\Gamma_{\rm en}$=64\,keV. Slowing down of  a 14.80\,MeV beam to 14.04\,MeV in a 14.5\,mg/cm$2$ carbon foil would produce $p_{\rm zz}=-0.19$ and $\Gamma_{\rm angle}=1.1\,^{\circ}$ and $\Gamma_{\rm en}$=79\,keV.

The curves of Fig.\,\ref{PzzProduction} are based on the cross section derived for the nominal areal target densities of Table\,\ref{Targets}. Due to the uncertainties, discussed in Sec.\,\ref{Uncertainties}, the achievable values of $p_{\rm zz}$ would be enhanced or reduced. Measurements with $E_{\rm in}$ in the gap between those with C94 and C129 would replace the smooth energy dependence between 14.0 and 14.8\,MeV by a more structured one.
\section{Summary~\label{Summary}}
The rapid fluctuations of the tensor polarization $p_{zz}$ of the deuteron beam behind the seven carbon targets, measured as a function of the initial beam energy $E_{\rm in}$ between 9.50 and 18.60\,MeV, are understood as fluctuations in the polarization production $\Delta p_{zz}(E)$ as a function of the deuteron energy $E$ in a 187.93\,mg/cm$^2$ carbon target in the slowing-down process from 18.60\,MeV to 9.50\,MeV. A satisfactory fit of $\Delta p_{zz}(E)/\Delta E$ is achieved with the use of 26 Gaussian-distributed cross sections $\sigma_{0}(E)$ describing removal of deuterons in the $m=0$ state from the beam ($\Delta p_{zz}(E)/\Delta E >0$) and 25 cross sections $\sigma_{\pm 1}(E)$ for removal of $m=\pm 1$ deuterons ($\Delta p_{zz}(E)/\Delta E <0$).\\*[-1mm]

Position and width of the (negative) peak in the coarse energy dependence of $p_{zz}$ correspond to excitation of the $^{14}$N giant dipole resonance. Below (above) the central excitation energy of 23.4\,MeV (deuteron energy 15.3\,MeV) removal of deuterons in the $m=\pm 1$ ($m=0$) state from the beam with the beam direction as quantization (z) axis prevails. This corresponds to dominance of $m=\pm 1$ oscillations of the deformed compound nucleus$^{14}$N along the major axis below 23.4\,MeV and $m=0$ oscillations along the minor axis above. The ratio of the mean oscillation energies below and above 15.3\,MeV agrees with the ratio from the hydrodynamic model.\\*[-1mm]

In the range of common energy between 9.5 and 16.9\,MeV the central deuteron energies $E_{0}$ from the present fit are found in excellent agreement with the energies of the narrow peaks in the excitation functions of the $^{12}{\rm C}(d,\alpha_{2})^{10}$B reaction. This reaction is understood as including intermediate excited $^{14}$N states. Due to the agreement in the resonance energies it is assumed that these states are created in the removal of deuterons from the beam, too. This assumption is supported by the similarity of the $^{14}$N excitation function, obtained from the present fit function, and other $^{14}$N excitation functions.\\*[3mm]
Strong evidence is found that the central energies $E_{0}$ of the fluctuation intervals with positive (negative) $\Delta p_{zz}(E)/\Delta E$ correspond to $^{14}$N states of positive (negative) parity resulting as intermediate states from an S-matrix $\chi ^{2}$ fit to the $^{12}{\rm C}(d,\alpha_{2})^{10}{\rm B}$ angular distributions. The parities of further $^{14}$N levels, established by other reactions, agree with the sign of $\Delta p_{zz}(E)/\Delta E$ in the energy range around these levels.\\*[-1mm]

Due to angular momentum and parity conservation,  in the $^{12}{\rm C}(d,\alpha_{2})^{10}$B reaction the orbital angular momenta in the initial an final state have to be equal, $l_{\rm i}=l_{\rm f}=l$, and spin and parity $J^{\pi}$ of the intermediate $^{14}$N states are confined to $J=l$ and $\pi=(-1)^{l}$. In the formation process $d+^{12}{\rm C}\rightarrow ^{14}{\rm N}^{*}$, however, excited states with $J=l\pm 1$ would be allowed as well as those with $J=l$. The states, resulting from the present fluctuation analysis, however, are in one-to-one agreement with the intermediate states in the $(d,\alpha_{2})$ reaction. This leads to the conclusion that essentially states with $J=l$ only are created in the removal of deuterons from the beam in the energy range of the present measurement.\\*[-1mm]

The 0.1\,MeV energy steps and the statistical errors of the present measurement do not allow to resolve narrow and weak peaks in the ranges of positive or negative $\Delta p_{zz}(E)/\Delta E$. The overall sign in the intervals, however, would be maintained.\\*[-1mm]

 The present results indicate the possibility to produce deuteron beams of tensor polarization $p_{\rm zz}$ between +0.3 and $-$0.4 by slowing down initially unpolarized deuteron beams with the use of carbon targets of appropriate areal density and a sandwich technique. The polarization values to be expected can be calculated with the use of the fit parameters of the Gaussian-shaped deuteron removal cross sections used to describe the fluctuations in $\Delta p_{zz}(E)/\Delta E$.\\*[-1mm]

The experimental setup of the present work had to be dismantled after the 2006 runs and thus no completing measurements could be performed (i) with higher statistical accuracy and energy steps $<$0.1\,MeV in selected energy intervals and (ii) to close the gaps in the initial beam energy of the present measurement. Such measurements were valuable with regard to the present conclusions from the fluctuation analysis.\\*[5mm]
{\bf Acknowledgements}  H.\,Seyfarth thanks Institut f\"{u}r Kernphysik for the hospitality. As a "Julumnus", a {\bf Ju}e{\bf l}ich al{\bf umnus} in the Julumni association, he thanks Forschungszentrum J\"ulich for the possibility to enter the campus. Thanks from A.~Rouba and V.~Baryshevsky go to Forschungszentrum J\"ulich for financial support of visits to J\"ulich. The help in the measurements by C.~D\"uweke, R.~Emmerich, A.~Imig, former members of Institut f\"{u}r Kernphysik, Universit\"at zu K\"oln, by M.~Mikirtychiants, former member of Institut f\"{u}r Kernphysik, Forschungszentrum J\"ulich, and by A.~Vasilyev from Petersburg Nuclear Physics Institute, Gatchina, Russia is thankfully acknowledged. The authors thank Zentralabteilung for Chemische Analysen  for the various analyses of the graphite foils, and Zentralbibliothek of Forschungszentrum J\"ulich for a literature search.\\*[5mm]


\begin{thebibliography}{999999}
\bibitem{ExpGra} Expanded graphite is made by immersing natural flake graphite in a bath of chromic acid, then concentrated sulfuric acid, which forces the crystal lattice planes apart, thus expanding the graphite.
\end{thebibliography}
\end{document}